\documentclass[a4paper,fleqn]{cas-sc}

\usepackage[round,authoryear]{natbib}
\usepackage{array}
\newcolumntype{C}[1]{>{\centering\arraybackslash}p{#1}}
\newcolumntype{L}[1]{>{\raggedright\arraybackslash}p{#1}}

\usepackage{enumitem}
\usepackage{adjustbox}
\usepackage{doi}  

\def\tsc#1{\csdef{#1}{\textsc{\lowercase{#1}}\xspace}}
\tsc{WGM}
\tsc{QE}

\begin{document}

\let\WriteBookmarks\relax
\def\floatpagepagefraction{1}
\def\textpagefraction{.001}

\shorttitle{Rotation period from combined datasets}

\shortauthors{K.Nagainis, I.Eglitis}

\title [mode = title]{Combination of multi-observatory photometry to determine asteroid rotation periods}

\author[1]{Kristers Nagainis}
\ead[url]{https://orcid.org/0009-0005-2475-9796}
\credit{Conceptualization, Methodology (combined dataset analysis), Software, Validation, Formal analysis, Investigation, Visualization, Writing -- review and editing, Project administration}

\affiliation[1]{organization={Institute of Astronomy, University of Latvia},
            addressline={Jelgavas str. 3},
            city={Riga},
            postcode={LV-1004},
            country={Latvia}}

\author[1]{Ilgmars Eglitis}
\date{\today}

\ead[url]{https://orcid.org/0000-0001-7652-815X }
\credit{Writing -- original draft, Methodology (separate dataset analysis), Data curation}

\begin{abstract}
This study aims to determine asteroid rotation periods from sparse, unevenly sampled, multi-observatory photometry using the Lomb-Scargle (L-S) method. The photometric measurements of the asteroids are taken from the circulars of the Minor Planet Center (MPC). Twelve asteroids, four of them with known rotation periods, were analyzed with two complementary methods. In the separate dataset analysis (S-method), which extends our earlier work, each observatory dataset was analyzed separately, and the final period was obtained as a weighted average of the consistent candidate periods, with weights based on the peak power of the power spectrum, the number of measurements, and the quality of the corresponding light curve fit. In the combined dataset analysis (C-method), all measurements were corrected for the biases of the photometric band, the star catalog and the observatory, reduced to a common scale and merged into a single dataset, after which the period was derived directly from the L-S analysis of the combined time series. Both methods produced consistent results, and they reproduced the known periods of the four test asteroids within 0.05\%. The results show that the L-S method is well suited for finding rotation periods in sparse observations with uneven sampling over long time intervals. Classical Fourier methods are less effective for such data. New rotation periods were found for asteroids 2968, 2971, 3081, 3473, 3716, and 4303. The period of 2968 remains ambiguous because of daily aliasing. For 3716, the most likely period is 10.474 hours, but its daily alias at 13.405 hours cannot be excluded completely. For 3173, the strongest peaks at 22.99 and 45.98 hours are probably daily aliases of a slower brightness variation. For asteroid 2607, the period agrees with a recently published value and is more precise. The approach is therefore promising for automated searches for unknown asteroid rotation periods.
\end{abstract}


\maketitle

\section{Introduction}\label{Introduction}

Asteroids pose a potential threat to Earth, as their impacts may have consequences ranging from local destruction to global extinction events \citep{Chapman1994, Shuvalov2013}. To mitigate such risks, it is essential to understand their physical properties in detail, including size, composition, structure, and rotational state \citep{Pravec2007, Warner2009, Durech2024}. The most direct approach is to study asteroids through dedicated spacecraft missions. However, such missions are costly and can only target a limited number of objects out of the millions that populate the Solar System. A more accessible alternative lies in the analysis of ground-based photometric observations, which provide a wealth of information about the rotation, shape, and surface characteristics of an asteroid \citep{Muinonen2010, Penttila2016, Mahlke2021, Oszkiewicz2023}. Nonetheless, it does provide only indirect constraints on asteroid physical properties and is, therefore, subject to possible ambiguity in the results.

Astro-photometric observations are primarily used for asteroid discovery and astrometric follow-up. Yet, when a sufficient number of measurements are obtained, they also reveal periodic brightness variations \citep{Warner2009, Veres2015, Rommel2025}, provided the rotation axis is not directed close to the line of sight. These variations arise from changes in the illuminated cross-section of an irregularly shaped, rotating body, leading to observable deviations from the brightness expected of a perfect sphere \citep{Pravec2007}. Therefore, photometric time series enable the determination of the rotational properties of an asteroid. Each asteroid exhibits a characteristic light curve, the brightness variation over one full rotation, which provides valuable information on both the spin period and the shape and elongation of the object \citep{Kaasalainen2001, Durech2015, Xu2025}. However, only the projection of the body along the line of sight is observed, and the amplitude and shape of the light curve change with the aspect angle, which differs between oppositions.

The most reliable light curves are obtained from continuous photometry of high precision. Asteroid observations, however, often have gaps in time coverage and are affected by changes in the atmosphere. Both add significant uncertainties to the measurements. To improve the analysis, more measurements are needed, but most asteroids in databases have insufficient observations at each observatory to obtain a reliable rotation period.

The Minor Planet Center (MPC) holds hundreds of millions of astrometric observations of asteroids, usually three or four per night for each object. Many of them also include the magnitude of the asteroid. The photometry is less precise than the astrometry, with a typical uncertainty of $\sim 0.1$ mag, and it contains systematic biases. These biases depend on the observatory and its instruments, on the choice of the star catalog, and on reference stars with a color different from that of the Sun \citep{Hoffmann2025}. The reported magnitudes also depend on the observing techniques, which have changed over time. The early measurements were made on photographic plates, while modern observations use charge-coupled device (CCD) cameras, which are much more accurate.

Combining many measurements in one photometric band, especially from several observatories, reduces the random errors. For each asteroid, we therefore combine the long-term observations of many observatories into one dataset, which covers many oppositions and a wide range of phase angles. The systematic biases are corrected as described in Section~\ref{sec: three}. Their possible time dependence, from gradual instrumental degradation or changes in the observational setup, is discussed in the same section.

Light curve analysis is a fundamental technique for the characterization of asteroids. It gives their rotation parameters, which provide critical insight into their internal structure, collisional history and dynamical evolution. Many asteroids with well-studied orbits, however, still lack measurements of such basic physical parameters. Of special interest are the near-Earth objects (NEOs) and asteroids with orbits subject to strong gravitational perturbations. Our sample continues that of \citet{Eglitis2025}, which consists of main-belt asteroids with an Earth Minimum Orbit Intersection Distance (MOID) \citep{Bowell1994, bonanno2000analytical} below 1.1 AU.

In this work, we analyze twelve asteroids with Earth MOID below 1.1 AU. Four of them have known rotation periods and are used to validate the methods. The data processing is explained in Section~\ref{sec: Data processing}. In Section~\ref{sec: Methods}, we describe the two methods, the separate dataset analysis (S-method) and the combined dataset analysis (C-method), and the Lomb-Scargle algorithm used in both of them. In Section~\ref{sec: results}, we present the results for all analyzed asteroids, while Section~\ref{sec: summary} summarizes the work. The validation with the four asteroids of known period is shown in Appendix~\ref{app: validation}, and the long-period analysis of (3173) McNaught in Appendix~\ref{app: 3173}.

\section{Data processing}\label{sec: Data processing}

For each asteroid, the photometry is taken from the MPC archive and processed with our own Python code \footnote{Code available here: \url{https://github.com/kristero/Combined-dataset-period-analysis/blob/main/mpc_catalog/fetch_mpc_obs.py}.}. The code first sorts the observations into datasets, each with the measurements of one observatory in one photometric band reported to the MPC. Datasets with fewer than 70 individual brightness measurements were excluded from further analysis. This threshold is a pragmatic choice. With $N = 70$, the $n = 2$ Fourier model of equation \ref{eq: fourier model} (five coefficients) still has 65 degrees of freedom. It must be noted that the reliability of a period also depends on the rotation period and light curve coverage, but this is unknown beforehand. Thus, the threshold is maintained as a simple filtering tool. 
A summary of the observational datasets utilized is provided in Table \ref{tab:observatories}.

\begin{table}[ht]
\centering
\caption{Observatories and surveys that provided the brightness data used in this work. Column 1 gives the IAU observatory code, Column 2 the observatory or survey project, Column 3 the MPC photometric band codes of the datasets, and Column 4 the total number of measurements of these datasets in Table~\ref{tab:asteroids}. The codes c and o denote the ATLAS cyan and orange bands. The code G denotes the Gaia G band and is also used for the TESS data (C57). TESS observes from space in a broad red band, and its large pixels make blending with stars possible. As for the other datasets, the constant magnitude offset of the TESS data is removed in both methods.}
\begin{tabular}{|c|l|l|r|}
\hline
\textbf{Code} & \textbf{Observatories/surveys} & \textbf{Bands} & \textbf{N} \\ \hline \hline
689 & U.S. Naval Observatory, Flagstaff & V & 145 \\ \hline
691 & Steward Observatory, Kitt Peak-Spacewatch & V & 74 \\ \hline
703 & Catalina Sky Survey & G, V & 4669 \\ \hline
704 & Lincoln Laboratory ETS, New Mexico (LINEAR) & V & 4486 \\ \hline
C57 & Transiting Exoplanet Survey Satellite & G & 12863 \\ \hline
D29 & Purple Mountain Observatory, XuYi Station & R & 633 \\ \hline
F51 & Pan-STARRS 1, Haleakala & w & 497 \\ \hline
F52 & Pan-STARRS 2, Haleakala & w & 77 \\ \hline
G45 & Space Surveillance Telescope, Atom Site & G, r & 1321 \\ \hline
G96 & Mt. Lemmon Survey & G, V & 1396 \\ \hline
H45 & Arkansas Sky Obs., Petit Jean Mountain South & R & 109 \\ \hline
I41 & Zwicky Transient Facility & g, r & 1285 \\ \hline
M22 & ATLAS South Africa, Sutherland & o & 1093 \\ \hline
P07 & Space Surveillance Telescope, HEH Station & G & 247 \\ \hline
T05 & ATLAS Haleakala & c, o, w & 5459 \\ \hline
T08 & ATLAS Mauna Loa & c, o, w & 5452 \\ \hline
W68 & ATLAS Chile, Rio Hurtado & c, o & 1794 \\ \hline
Y00 & SONEAR Observatory, Oliveira & R & 99 \\ \hline
\end{tabular}
\label{tab:observatories}
\end{table}

\begin{table}[ht]
\centering
\caption{Summary of the asteroid brightness data retrieved from the MPC for this study. Columns 1-3 identify the asteroid by number, name and designation, and Column 4 gives its Earth MOID. Column 5 lists the datasets with at least 70 measurements by IAU observatory code and photometric band, with their numbers of brightness measurements in parentheses. Column 6 gives the sum of these numbers. Columns 7 and 8 give the time span (yyyy.mm) and the phase angle range (degrees) of these datasets. The datasets used in the C-method and their numbers of measurements after outlier removal are listed in Table~\ref{tab:combined}.}
\scriptsize
\setlength{\tabcolsep}{2pt}
\renewcommand{\arraystretch}{0.9}

\resizebox{\textwidth}{!}{%
\begin{tabular}{|C{0.06\linewidth}|C{0.085\linewidth}|C{0.1\linewidth}|C{0.06\linewidth}|L{0.35\linewidth}|C{0.06\linewidth}|C{0.06\linewidth}|C{0.06\linewidth}|}
\hline
\textbf{Number}&{\textbf{Name}}&{\textbf{Designation}} & \textbf{Earth MOID (AU)} & \textbf{Observatory codes and photometric bands} & \textbf{Number of obs.} & \textbf{Obs. date interval} & \textbf{Phase interval (deg)} \\ \hline
1951 & Lick & 1949 OA & 0.30579 & 703G(228), 703V(87), C57G(1753), H45R(109), I41g(92), I41r(106), M22o(111), T05c(103), T05o(220), T05w(84), T08c(76), T08o(295), W68o(175), Y00R(99) & 3538 & 2006.02-2024.01 & 13.6-51.1 \\ \hline
1963 & Bezovec & 1975 CB & 0.93243 & 689V(145), 703G(156), 703V(174), 704V(70), C57G(614), I41r(115), M22o(194), T05c(258), T05o(386), T05w(122), T08o(634), T08w(79), W68c(87), W68o(245) & 3279 & 1999.06-2024.01 & 3.4-31.2 \\ \hline
2134 & Dennispalm & 1976 YB & 1.0979 & 703G(215), 703V(75), 704V(260), C57G(119), G45r(144), I41g(107), I41r(92), T05c(231), T05o(229), T08o(292), W68o(96) & 1860 & 1999.03-2024.01 & 2.8-30.5 \\ \hline
2150 & Nyctimene & 1977 TA & 0.84457 & 703G(275), 703V(135), 704V(318), C57G(517), G45r(142), I41r(134), M22o(79), T05c(174), T05o(341), T05w(88), T08c(90), T08o(526), W68o(127) & 2946 & 1998.08-2024.01 & 1.5-33.6 \\ \hline
2607 & Yakutia & 1977 NR & 0.82911 & 703G(149), 703V(198), 704V(534), D29R(81), F51w(73), G45r(194), G96G(113), G96V(93), T05c(134), T05o(323), T08o(377) & 2269 & 1998.03-2024.01 & 0.4-32.4 \\ \hline
2968 & Iliya & 1978 QJ & 0.62344 & 703G(111), 703V(114), 704V(262), F51w(77), G45G(90), G45r(109), G96G(75), T05c(81), T05o(139), T08c(76), T08o(297) & 1431 & 1999.02-2024.02 & 0.4-37.1 \\ \hline
2971 & Mohr & 1980 YL & 0.99639 & 703G(278), 703V(173), 704V(499), C57G(1972), F51w(87), G45r(157), G96G(101), I41r(156), M22o(115), T05c(205), T05o(278), T08o(501), W68o(218) & 4740 & 1999.05-2023.12 & 0.1-30.3 \\ \hline
3081 & Martinuboh & 1971 UP & 0.96276 & 703G(174), 703V(182), 704V(487), C57G(694), D29R(95), G96G(95), G96V(99), I41r(74), M22o(87), T05c(123), T05o(97), T08o(232), W68o(225) & 2664 & 1999.02-2024.01 & 0.5-29.9 \\ \hline
3173 & McNaught & 1981 WY & 0.73088 & 703G(408), 703V(196), 704V(522), C57G(2901), D29R(90), F51w(77), G45r(143), G96G(110), I41r(132), M22o(126), T05c(211), T05o(258), T05w(76), T08o(526), W68o(180) & 5956 & 1999.02-2024.01 & 2.2-34.6 \\ \hline
3473 & Sapporo & A924 EG & 0.99465 & 691V(74), 703G(168), 703V(201), 704V(531), C57G(1817), D29R(96), G45r(91), G96G(93), G96V(166), I41r(277), M22o(85), T05c(159), T05o(162), T08o(386), W68o(153) & 4459 & 1991.09-2023.12 & 0.4-30.7 \\ \hline
3716 & Petzval & 1980 TG & 0.86715 & 703G(220), 703V(229), 704V(560), C57G(124), D29R(144), F51w(73), F52w(77), G45r(123), G96G(128), G96V(119), M22o(110), P07G(80), T05c(146), T05o(373), T08o(531), W68o(86) & 3123 & 1998.04-2024.06 & 0.4-32.2 \\ \hline
4303 & Savitskij & 1973 SZ3 & 0.89591 & 703G(292), 703V(231), 704V(443), C57G(2352), D29R(127), F51w(110), G45r(128), G96G(128), G96V(76), M22o(186), P07G(167), T05c(161), T05o(297), T08o(534), W68o(202) & 5434 & 1999.10-2025.02 & 0.4-31.9 \\ \hline
\end{tabular}%
}
\label{tab:asteroids}
\end{table}

An overview of the asteroids included in this study is presented in Table \ref{tab:asteroids}.
Apart from rotation, the magnitude of an asteroid depends on its heliocentric and geocentric distances and on the phase angle. The magnitudes were reduced to 1 AU from the Sun and the observer, $m - 5\log_{10}(r\Delta)$, where $r$ and $\Delta$ are the heliocentric and geocentric distances in AU. The observation times were corrected for the light-travel time, $t = t_{\mathrm{obs}} - \Delta/c$, where $t_{\mathrm{obs}}$ is the time of the observation and $c$ is the speed of light. Both corrections were made in the same way as in \citet{Eglitis2023}. The phase angle dependence is removed as part of the methods of Section~\ref{sec: Methods}. The corrected magnitudes of each dataset plotted against the phase angle give a phase diagram for each dataset of each asteroid.

Outliers are then removed from these diagrams by sigma clipping in $5^\circ$ bins of phase angle. A measurement is removed if it deviates from the mean magnitude of its bin by more than three standard deviations of the magnitudes in the bin \citep{ivezic2020statistics}. Below $7^\circ$ phase angle, the limit is $5\sigma$, so that the opposition surge is not removed. Each bin contains all rotational phases, so the rotational variation is not removed as a whole. This clipping is applied to every dataset in both methods, and the C-method uses the further limits of step 2 of Section \ref{sec: three}.

\section{Methods} \label{sec: Methods}
\subsection{Combining brightness data} \label{sec: three}

Two methods were employed to obtain the rotation period from the brightness data of many observatories. The first, the separate dataset analysis (S-method), was introduced by \citet{Eglitis2025} and is summarized in Section~\ref{subsec: seperate dataset}. In the S-method, each dataset is analyzed on its own. The corrections and the outlier removal of Section~\ref{sec: Data processing} are the same for both methods.

In the second method, the combined dataset analysis (C-method)\footnote{The code is available here: \url{https://github.com/kristero/Combined-dataset-period-analysis}.}, all asteroid observations were reduced to 1 AU, following \citet{Mahlke2021}, allowing all series to be combined into a single consistent dataset. There are several advantages of combining observations from different observatories and analyzing their rotation curves, in comparison to analyzing different datasets separately, and only then combining their results.

Firstly, the temporal coverage is better, which reduces the noise of the Lomb-Scargle power spectra. Additionally, it reduces aliasing from window function effects, as there are fewer gaps due to observational restrictions from a single observatory. For example, if a single observatory observed an asteroid every day, there would be a gap of approximately 24 hours, which would, in turn, give a false period of 24 hours. The temporal coverage throughout the day reduces aliasing, thus allowing for a better determination of the physical rotation period.

Aliasing is used throughout Section~\ref{sec: results} to judge the candidate periods, so it is described here in more detail. The window function of ground-based data has peaks at multiples of one cycle per sidereal day ($1.0027$ days$^{-1}$), which come from the nightly gaps, and at multiples of one cycle per year, which come from the gaps between the oppositions. A double-peaked light curve with the period $P$ (in hours) varies with the frequency $\nu = 48/P$ days$^{-1}$, which is twice the rotation frequency $f = 24/P$ days$^{-1}$ of equation \ref{eq: fourier model}. Its daily aliases appear at the frequencies $|\nu - k \cdot 1.0027|$ days$^{-1}$, where $k$ is an integer. Two periods are therefore daily aliases of each other if $48/P_1 - 48/P_2 \approx k$ days$^{-1}$. For slow rotators, the alias frequency can be reflected around zero, and the two periods then satisfy $48/P_1 + 48/P_2 \approx k$ days$^{-1}$ instead. The yearly gaps produce a fine comb of peaks around every peak, separated by about $1/365$ days$^{-1}$ in $\nu$, which corresponds to the uncertainty in the number of rotations between the oppositions. Peaks at 24, 12 and 8 hours are produced directly by the window function, for example when the magnitude offsets between nights or observatories are not removed completely. In addition, the half period $P/2$ always gives a peak of similar power, because the light curve folded with $P/2$ is close to a single sine wave (Section~\ref{Yakutia}). Combining observatories at different longitudes weakens the daily aliases, but it does not remove them. For this reason, the daily aliases of every adopted period are identified and discussed in Section~\ref{sec: results}.

The second advantage of combining the observations is that the total time baseline ($\Delta T$) increases substantially, which in turn increases the frequency resolution ($\sim 1/\Delta T$) for the possible rotational curves. Similarly, the signal-to-noise ratio (SNR) increases as $\sim \sqrt{N}$, where $N$ is the total number of measurements. Additionally, the phase coverage is better, which is important for asteroids with longer rotation periods. The larger baseline also improves the accuracy of the period and reduces spurious periods. The longest period that can be found is limited by the coverage within one apparition and by slow systematic variations, not by $\Delta T$. Periods longer than 240 hours were therefore searched only for 3173 (Appendix~\ref{app: 3173}).

The magnitudes reported to the MPC have systematic offsets that depend on the photometric band, on the star catalog used in the reduction and on the observatory \citep{Hoffmann2025}. In this work, every measurement is corrected with the DePhOCUS offsets of \citet{Hoffmann2025}, which convert the reported magnitude to the Johnson V band. The offset is taken from the most suitable rules available for the combination of observatory, band and star catalog, and otherwise from the band alone. We used the corrections of the confidence level 0.90 recommended by \citet{Hoffmann2025}. The DePhOCUS offsets of the bands alone agree with our own analysis of the band offsets\footnote{Our own analysis of the offsets of the bands and observatories is available here: \href{https://github.com/kristero/Asteroid_abs_mag/blob/main/All_asteroid_iteration.ipynb}{GitHub bias analysis}}. With the DePhOCUS corrections, the magnitudes of all datasets are on the V scale, and the absolute magnitudes of the asteroids can be estimated from all datasets together (Table~\ref{tab:absmag}). The rotation periods are the same with and without the DePhOCUS corrections (differences below $0.0001$ hours), because the constant offsets are also removed in step 5 of the pipeline below.

Additionally, the evolving instrumental setups can affect the brightness measurements in a time dependent manner. This can be due to changes in instrumental setup, deteriorating quality of the instruments, or even changes in data post-processing. Part of these changes is removed by the DePhOCUS corrections, because the star catalog used by an observatory changed over time. For example, the corrections of the 703G measurements reduced with the Gaia DR1 and Gaia DR2 catalogs differ by $0.32$ mag. The remaining changes are not removed by the constant magnitude shift of step 5, so they remain in the combined data. Each observatory covers only some nights and oppositions, so these offsets do not only add scatter. They can also add power at long periods and at their daily aliases. For asteroids 3173 and 4303, their effect on the period was checked by analyzing the oppositions separately (Sections~\ref{subsec: 3173} and \ref{subsec: 4303}).

The pipeline of the developed method is as follows:
\begin{enumerate}[label=\arabic*)]
    \item \textit{Choosing the observatories}. After applying the corrections, each observatory's data is inspected, because not all data can be used. The procedure is described in more detail in the last step. The reference observatory should cover phase angles below $7^\circ$, which are essential for the $G_1$ parameter, and a wide range of larger phase angles. It has been found that the Mauna Loa (T08o) observatory fulfills these requirements for most asteroids. For consistency, it is taken as the reference observatory in all cases (Table \ref{tab:combined}). If the data of a particular observatory do not align well with the other observatory data, its measurements are discarded.

    \item \textit{Outlier removal}. Each observatory's dataset still contains measurement outliers, most commonly in the first or last observations of the night. These points are typically obtained at dusk or dawn, when observing conditions are less stable, and may therefore be affected by imperfect conditions or other sources of measurement inaccuracy. The outliers are identified by sigma clipping in bins of $5^\circ$ of phase angle (Section~\ref{sec: Data processing}). For the reference fit (step 3), measurements that deviate from the mean of their bin by more than $3\sigma$ are removed. Before the combination (steps 4 and 5), a stricter limit of $1.8\sigma$ is applied to all datasets. Below $7^\circ$ phase angle, the limit is $5\sigma$, so that the opposition surge is not removed. After the removal of the phase dependence (step 6), the remaining outliers are identified by finding the interquartile range and applying Tukey's rule (with a factor of 1.8) to discard measurements that are outside an approximately $3\sigma$ range.

    \item \textit{$HG_1G_2$ computation for reference observatory}. Each asteroid has its phase curve, which is the brightness variation with respect to the Solar phase angle. It can be described by different models, but in this work, the $H G_1 G_2$ description by \citet{Penttila2016} is used:

    \begin{equation}
    V(\alpha) = H-2.5\log_{10}\left[G_1\Phi_1(\alpha)+G_2\Phi_2(\alpha)+(1-G_1-G_2)\Phi_3(\alpha)\right]
    \end {equation}
     where $H$ is the absolute magnitude, corresponding to the brightness the asteroid would have if observed at heliocentric and geocentric distances of 1 AU and at zero phase angle ($\alpha = 0^\circ$). $G_1$ and $G_2$ are empirical slope parameters that describe the relative contributions of the three basis phase functions $\Phi_1(\alpha)$, $\Phi_2(\alpha)$, and $\Phi_3(\alpha)$. These coefficients determine the shape of the phase curve, i.e., how the brightness decreases with increasing phase angle. $\Phi_1(\alpha)$, $\Phi_2(\alpha)$, and $\Phi_3(\alpha)$ are the standard basis functions as defined in \citet{Muinonen2010}.

     This model is fitted to reference observatory data by least squares with a robust loss function and the constraints $G_1 \geq 0$, $G_2 \geq 0$ and $G_1 + G_2 \leq 1$, obtaining the $H$, $G_1$, and $G_2$ values (Table \ref{tab:combined}). The MPC reports do not contain individual brightness uncertainties. Therefore, the fit is unweighted, and the formal errors of the parameters are indicative only. Because the magnitudes are corrected to the V band, $H$ is the absolute magnitude in V obtained from the reference observatory.

    \item \textit{$HG_1G_2$ computation for the rest of the observatories}. The fit of step 3 is repeated for the other observatories, but now the $G_1$ and $G_2$ parameters are fixed to the reference values, and only the $H$ is left free. This means that the fitted curve for the reference observatory is simply shifted in magnitude for the other observatories, irrespective of phase difference. The premise of the method is that, to first approximation, the $G_1$ and $G_2$ parameters for an asteroid remain unchanged, regardless of the bands used. \footnote{Wavelength dependent differences in the phase curve were found by \citet{Mahlke2021} and \citet{Colazo2025}, and for the opposition effect by \citet{Slyusarev2025}. To test their effect, $G_1$ and $G_2$ were also fitted separately for every band with at least 150 measurements that cover phase angles below $10^\circ$ and a range of at least $15^\circ$, after the magnitude offsets of the datasets of that band were removed. The periods of all asteroids were unchanged (differences below $0.0001$ hours). However, the slope parameters of the individual bands were poorly constrained by the MPC data, and many of them reached the limits of the constraints. Therefore, the reference values are used for all bands, and the remaining wavelength dependent differences are part of the scatter of the combined data.}

    \item \textit{Shifting to the reference fit}. After fitting, each dataset $j$ is shifted to match the reference observatory phase curve. This is done by computing $\Delta H_j = H_j - H_{\mathrm{ref}}$ between the observatory of interest and the reference observatory, where $H_j$ and $H_{\mathrm{ref}}$ are the values fitted in steps 4 and 3. This magnitude shift is then applied to the corrected magnitudes $m_j$ of the dataset, $m_{\mathrm{shifted}} = m_j - \Delta H_j$. It removes the constant offsets between the datasets that remain after the DePhOCUS corrections. For the asteroids of this work, the shifts are between $-0.27$ and $0.33$ mag, with an rms of $0.03$-$0.13$ mag for each asteroid. This yields a unified dataset with the same profile as a function of phase angle, regardless of the observatory or band.

    \item \textit{Phase dependence removal}. The phase dependence is removed to fit the light curves.
    After shifting the observations from the different observatories onto a common magnitude scale,
     $m_{\mathrm{shifted}}(\alpha)$ denotes the shifted observed magnitude at phase angle $\alpha$,
    and $V_{\mathrm{fit}}(\alpha)$ is the fitted phase function model for these shifted data.
    If $H$ is the phase independent absolute magnitude level of the fitted model, then the phase dependent contribution is

    \begin{equation}
    \Delta(\alpha) = V_{\mathrm{fit}}(\alpha) - H .
    \end{equation}

    The phase reduced magnitude used for the rotation period analysis is then obtained by subtracting this phase term from the shifted observations:

    \begin{equation}
    H_{\mathrm{reduced}} = m_{\mathrm{shifted}}(\alpha) - \Delta(\alpha).
    \end{equation}

    Thus, $H_{\mathrm{reduced}}$ represents the magnitude variations after removal of the phase angle dependence, and it is the quantity used for the rotational period computation.

    \item \textit{Revision of observatory data}. After obtaining the phase curve of the combined dataset, the observatory datasets are reviewed again to decide whether all of them should be retained. Three criteria for data exclusion are checked when the $H G_1 G_2$ model is fitted to the phase curve of each dataset. (1) The best-fit parameters of a free fit converge to the boundary values $G_1 = 0$ and $G_2 = 1$ (or vice versa). (2) The reduced $\chi^2$ of the fit with fixed $G_1$ and $G_2$ exceeds $3\chi_{\mathrm{ref}}^2$, where $\chi_{\mathrm{ref}}^2$ is the reduced $\chi^2$ of the same fit for the reference observatory. (3) The data show clearly unphysical patterns (most commonly in survey data, for example measurements of the same night that differ by about one magnitude or more). A dataset is discarded if criterion (1) or (3) applies.  Criterion (2) marks a dataset for a closer check, in which the C-method (Section~\ref{subsec: combined}) is computed with and without the dataset. The dataset is kept if it does not lower the power $w$ of the period peak, that is, if it does not reduce the quality of the light curve fit.
\end{enumerate}

\begin{figure}
    \centering
    \includegraphics[width=1.\linewidth]{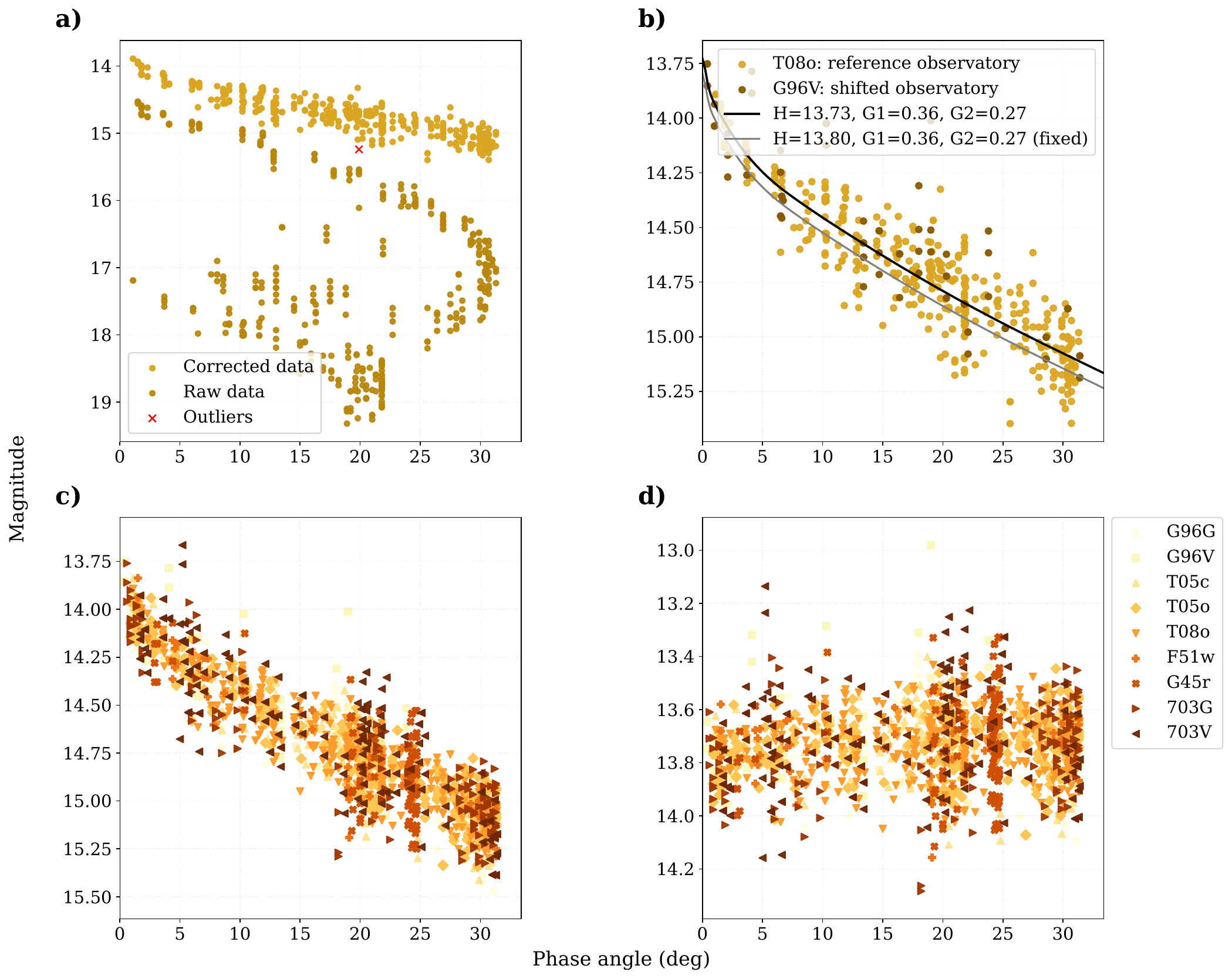}
    \caption{Pipeline visualization for combining data from different observatories for asteroid 2607. (a) Distance, light-time and DePhOCUS corrections, and outlier removal (crosses) for the reference observatory T08o (steps 1-2). (b) Fit the $H G_1 G_2$ model to the reference phase curve (black line), fit the G96V data with fixed $G_1$ and $G_2$ (gray line), and shift them to the reference fit (steps 3-5). (c) Combine all retained datasets (steps 5 and 7). (d) Shift the data to zero phase angle (step 6). The reduced magnitudes scatter around the level $H$ of the reference fit.}
    \label{fig: Pipeline for mag}
\end{figure}

The end product of this pipeline is combined data from different observatories, independent of phase angle. For clarity, the pipeline is visualized in Fig. \ref{fig: Pipeline for mag}.

\subsection{Lomb-Scargle method}\label{sec: L-S}

Because of its irregular, elongated shape, a rotating asteroid shows periodic brightness changes, and its rotation period can be derived from them. Continuous monitoring, however, is often not possible. Most of our measurements come from ground-based telescopes, whose observations are interrupted by the day and night cycle and by the weather. The light curves therefore have large gaps, and the periodicity is not always obvious.

We use the Lomb-Scargle method \citep{Lomb1976, Scargle1982, VanderPlas2018} to find rotation periods in such incomplete data. It works well for sparse time series with uneven or irregular sampling, and it is fast enough for large datasets. Because it is based on a least-squares fit of a periodic model, it also separates true periodic signals well from white noise. For these reasons, it is our preferred method for the period analysis.

Both the power spectra and the model light curves are computed with the Lomb-Scargle implementation of Astropy \citep[version 6.1,][]{astropy:2022}. At every trial frequency $f$, the method fits a truncated Fourier series to the measurements:
\begin{equation}
\label{eq: fourier model}
m(t) = c + \sum_{k=1}^{n} \left[ a_k \sin(2\pi k f t) + b_k \cos(2\pi k f t) \right],
\end{equation}
where $c$, $a_k$ and $b_k$ are the fitted coefficients and $n$ is the number of harmonics. The power $w$ is the relative reduction of $\chi^2$ with respect to a constant model. In this work, $n=2$ is used for the period search, as it is the lowest order that reproduces the two maxima and two minima of an elongated body. The period search covers $0.5$-$240$ hours ($0.1$-$48$ days$^{-1}$) in both methods. This range follows the expected distribution of asteroid rotation periods \citep{Pravec2007}. It encompasses the limit of $2.2$ hours for rubble pile asteroids and also includes the rare case of faster rotators. Slow rotators are also included, while the range remains compatible with the typical temporal baseline and sampling of the available photometric data, beyond which period determination becomes increasingly degenerate and unreliable.
In the C-method, the frequency grid is sampled with 15 points per peak width $1/\Delta T$.
For each power spectrum, the threshold for significant peaks is $5 \sigma$ above the mean power. In the power spectrum figures of Section \ref{sec: results}, the peaks above the threshold given in the legend are marked with their periods. Where a few peaks dominate the power spectrum, the threshold for marking peaks in the figure was raised above the $5 \sigma$ level, so that only these peaks and their strongest alternatives are labeled.

The light curves are folded with the rotational phase, which is the fractional part of $(t - t_0) f$ expressed in degrees. Here $t$ is the light-time corrected Julian date, $f$ is the rotation frequency and $t_0$ is the zero-phase epoch, taken as the first observation of the analyzed dataset. The value of $t_0$ is given in the figure captions. Phase zero is therefore an arbitrary reference point and is not related to the Solar phase angle. The light curve figures show the $n=4$ model of equation \ref{eq: fourier model} at the period of the $n=2$ power spectrum peak, because it follows the shape of the light curve in more detail.

It must be noted that there cannot be absolute certainty about the obtained rotation periods. All of them are subject to the limitations of the sampling, where two periods cannot be distinguished if $|P_1 - P_2| \lesssim P^2/\Delta T$, where $P$ is the true (or candidate) period and $\Delta T$ is the total time baseline. Another limitation is aliasing (Section~\ref{sec: three}), which affects both methods. In a single dataset, the daily aliases can be as strong as the period itself. Combining the datasets weakens them, but it does not remove them, and a high value of the period significance criterion of Section~\ref{subsec: seperate dataset} does not exclude an alias. The orientation of the spin axis is not known, so the derived periods are synodic. The synodic period differs from the sidereal one by about $\omega P^2/(360^\circ \cdot 24)$ hours, where $\omega$ is the rate of change of the longitude of the phase angle bisector in degrees per day and $P$ is in hours. The epoch and shape of the light curve also change between oppositions, which limits the coherent folding of the combined dataset. Additionally, the asteroids with a small amplitude of the light curve or low SNR have less reliable rotation periods, which must be kept in mind in the analysis. No lower limit on the amplitude was imposed.

The S-method and the C-method derive the period separately, although the strongest peaks of the C-method are also used as candidates in the S-method (Section~\ref{subsec: seperate dataset}). Each method has its advantages and shortcomings, which will be explained in the following sections.
\subsection{Separate dataset analysis (S-method)} \label{subsec: seperate dataset}

In the S-method, every dataset of Table~\ref{tab:asteroids} is analyzed separately. The datasets that are not used in the C-method (notes of Table~\ref{tab:combined}) are included, and the criteria below decide whether they are used. The phase angle dependence of each dataset is removed with a free $HG_1G_2$ fit to that dataset alone (step 3 of Section~\ref{sec: three}). The Lomb-Scargle power spectrum ($n=2$) is computed over the period range of $0.5$-$240$ hours ($0.1$-$48$ days$^{-1}$) given in Section~\ref{sec: L-S}, with 10 points per peak width. The candidate periods are the strongest peaks of the largest datasets and of the C-method. For each candidate, the power spectrum of every dataset is computed again within $\pm 3\%$ of the candidate frequency, with 50 points per peak width, and the highest peak in this interval gives the period $p_i$ and its power $w_i$ of the dataset.

\begin{figure}
    \centering
    \includegraphics[width=0.5\linewidth]{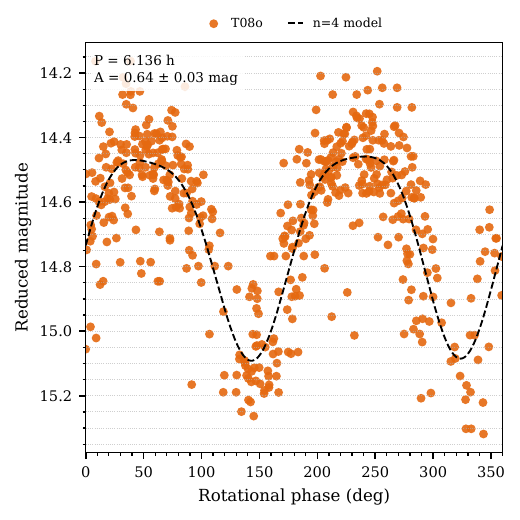}
    \caption{The light curve of asteroid 4303 obtained from T08o data, folded with the period of $6.136$ hours (phase 0 at JD 2457926.88). The dashed line is the $n=4$ Fourier model, and the amplitude $A$ is measured as described in Section~\ref{subsec: amplitude}.}
    \label{fig: Light curve 4303 T08o}
\end{figure}

\begin{figure}
    \centering
    \includegraphics[width=0.9\linewidth]{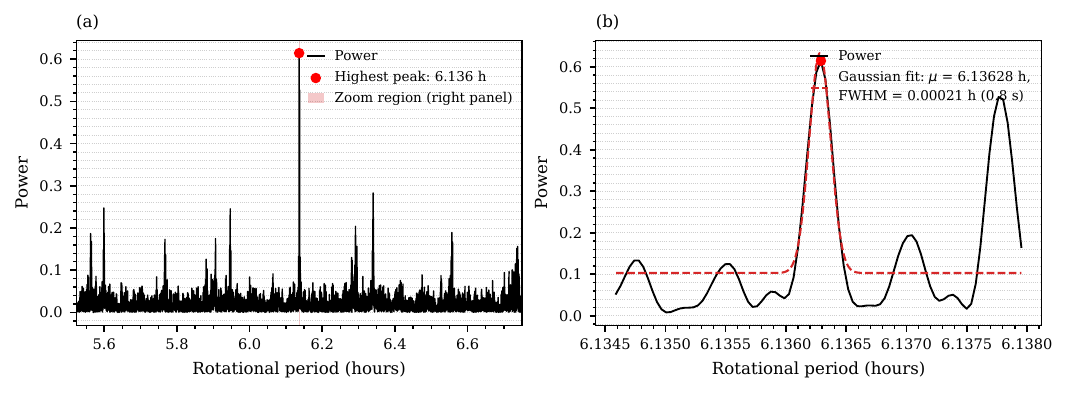}
    \caption{The power spectrum of asteroid 4303 obtained from T08o data. (a) Within $10\%$ of the highest peak. (b) Zoom on the highest peak at $P = 6.136$ hours with the fitted Gaussian profile (dashed line), which has its center at $6.13628$ hours and a full width at half maximum of $0.00021$ hours. The other peaks in panel (b) are the yearly aliases of the period.}
    \label{fig: Power spectrum 4303 T08o}
\end{figure}

A dataset is used for a candidate if it meets three criteria. The first two were introduced by \citet{Eglitis2023}. (1) The light curve folded with $p_i$ has two maxima and two minima, which is the typical shape of an elongated rotating asteroid (Fig. \ref{fig: Light curve 4303 T08o}). (2) The peak is at least $5\sigma$ above the mean power of the dataset, where $\sigma$ is the standard deviation of its power spectrum. Such peaks have a nearly Gaussian profile (Fig. \ref{fig: Power spectrum 4303 T08o}). (3) The period $p_i$ is consistent with the other datasets. The dataset is rejected if its $p_i$ deviates from the median $p_i$ of the datasets that meet criteria (1) and (2) by more than $3s$, where $s$ is the larger of the robust standard deviation ($1.4826$ times the median absolute deviation) and the spacing of the yearly aliases, $P^2/(24 \cdot 730.5)$ hours. The double-peaked shape is the most common one, but it is not universal, because asteroids with strongly asymmetric shapes can show one or more than two maxima per rotation \citep{Harris2014}. For this reason, the light curves are also shown with the higher-order $n=4$ Fourier model (Section \ref{subsec: combined}).

The rotation period of the S-method is the weighted average of the periods of all datasets that meet these criteria. Following \citet{Eglitis2025}, the weights are based on three quantities of each dataset, namely the peak power, the number of measurements, and the coefficient of determination $R_i^2$ of an $n=4$ Fourier fit of the light curve folded with $p_i$:

\begin{equation}
P = \frac{
\left( \sum_i \left( \frac{w_i}{w_s} p_i \right) + \sum_i \left( \frac{N_i}{N_s} p_i \right) + \sum_i \left( \frac{R_i^2}{R_s^2} p_i \right) \right)
}{3}
\end{equation}

where $P$ is the final period of the asteroid. For the $i$-th dataset, $p_i$ is the period, $w_i$ the peak power, $N_i$ the number of measurements and $R_i^2$ the coefficient of determination of its light curve. The sums of $w_i$, $N_i$ and $R_i^2$ over all used datasets are $w_s$, $N_s$ and $R_s^2$. The resulting period is shown in Column 3 of Table \ref{tab:results}.
The uncertainty of $P$ is estimated with the square weighted mean period error $\sigma_{kv}$ \citep{Eglitis2025}:
\begin{equation}
\label{eq: sigma kv}
\sigma_{kv} = \sqrt{\frac{\sum_{i=1}^{M} (p_i - P)^2}{(M - 1)M}}
\end{equation}

where $M$ is the number of datasets in the average. This error does not include the probability that the true period is an entirely different one, for example, the period of the second highest peak.
In cases where multiple candidate rotation periods are identified, an additional parameter, the \textit{period significance criterion}, is used. It requires that the found period has simultaneously a large power spectrum peak value and a small dispersion of the periods obtained from the different datasets. It is given by equation \ref{eq: period significance}.

\begin{equation}
\label{eq: period significance}
\xi = \frac{W_{av}\, P}{\sigma_{kv}}, \qquad \log \xi \equiv \log_{10}\frac{\xi}{50},
\end{equation}

where $W_{av}$ is the average of power spectrum peak values, $\sigma_{kv}$ (see equation \ref{eq: sigma kv}) is the period mean square error, and $P$ is the weighted mean period. The ratio $P/\sigma_{kv}$ is the inverse of the relative dispersion of the period, so $\xi$ increases with the peak power and with the agreement between the datasets. The number of measurements does not enter $\xi$ directly. If there are multiple possible solutions for the period, the one with the highest $\log\xi$ is the most probable true rotation period of the S-method. When several candidates are daily aliases of each other and their $\log\xi$ values are all close to the threshold or all well above it, the S-method alone cannot decide between them, and the power spectrum of the C-method is also used. For these asteroids, Table~\ref{tab:results} gives the S-method candidate that agrees with the C-method period, and the other candidates are given in the notes of the table. The threshold $\xi = 50$ is empirical. It was set from our earlier analyses of asteroids with multiple candidate periods, and periods with lower $\xi$ are not considered reliable. Such uncertainty may arise from multiple periodicities caused by the precession of the rotation axis, or from a significant obliquity, where the spin axis is strongly tilted relative to the orbital plane. Since $\xi$ can span a wide range (from single-digit values up to tens of thousands), we use the logarithmic scale $\log\xi = \log_{10}(\xi/50)$ of equation \ref{eq: period significance}, so that the threshold $\xi = 50$ corresponds to $\log\xi = 0$. This implies that periods with $\log\xi < 0$ are likely spurious, but those with $\log\xi > 0$ are more likely to be the true periods. Larger values of $\xi$ correspond to higher confidence in the derived period. For example, the S-method gives $W_{av} = 0.24$, $P = 2.918$ hours and $\sigma_{kv} = 0.0112$ hours for asteroid 2607, which give $\xi = 62$ and $\log\xi = 0.09$, just above the threshold.

\subsection{Combined dataset analysis (C-method)} \label{subsec: combined}

In the C-method, the datasets from different observatories were combined following the pipeline described in Section \ref{sec: three}. The Lomb-Scargle method was applied for the same period range of 0.5-240 hours (0.1-48 days$^{-1}$) as in the S-method. The abundance of measurements, compared to the S-method, reduces the noise in the power spectra and provides more accurate periods due to increased frequency resolution.
This is because even a slight difference in period can produce drastically different light curves, as the long time baseline causes errors to accumulate.

The C-method analyzes only one combined dataset, so it does not need the second search within $\pm 3\%$ of each candidate frequency that is used in the S-method. A higher-order Fourier model ($n=4$) can approximate the light curves with greater detail, but it also increases noise and aliasing. Therefore, $n=2$ is used for the period search and for all power spectra shown in this work. There is no universal quantitative rule for the number of harmonics without knowledge of the shape of the asteroid. The $n=4$ model is used for the light curve figures and for the amplitudes (Section~\ref{subsec: amplitude}).

The Lomb-Scargle method gives no formal error of the period. As an estimate of the uncertainty of the C-method period, we take the period resolution $P^2/\Delta T$ (Section \ref{sec: L-S}), which is given in Column 7 of Table~\ref{tab:results}. It is below $0.002$ hours for all asteroids except 3173 ($0.012$ hours). It is not an upper limit of the error, because the periods of the two methods, and the synodic and sidereal periods (Section \ref{sec: L-S}), can differ by more than the period resolution. This estimate also does not include the possibility that the true period is an alias, so the alternative periods of each asteroid are discussed in Section \ref{sec: results}. In the future, Markov Chain Monte Carlo (MCMC) methods can be used to estimate the uncertainties.

To check the effect of the changing aspect and of the time dependent offsets, the combined data of 3173 and 4303 were also divided into oppositions, using the opposition dates of the JPL Horizons ephemerides. Each opposition with at least 50 measurements was analyzed with the same Lomb-Scargle settings.

\subsection{Light curve amplitude} \label{subsec: amplitude}

The amplitude of a combined light curve is reduced when the light curve changes between the apparitions, because the viewing geometry and the phase of the extrema change \citep{Waszczak2015}. The peak-to-peak range of the $n=4$ model of the combined data is therefore not used as the amplitude. For the four test asteroids (1951, 1963, 2134, 2150), it is $0.16$, $0.26$, $0.17$ and $0.54$ mag, which is $13$-$47\%$ below the medians of their published amplitudes (Appendix~\ref{app: validation}). Instead, the amplitude is measured for every apparition separately, following the single apparition Fourier fits of survey photometry \citep{Waszczak2015, Chang2015}. The data are divided into apparitions at gaps longer than 60 days, and an apparition is used if it contains at least 60 measurements that cover at least 8 of 10 intervals of rotational phase as an emperical choice. A common light curve shape $S(\phi)$, an $n=4$ Fourier series of the rotational phase $\phi$ with unit peak-to-peak amplitude, is fitted to every apparition $j$ as
\begin{equation}
m_i = Z_{d,j} + g_j\, S(\phi_i - \Delta\phi_j),
\end{equation}
where $Z_{d,j}$ is a zero point of the dataset $d$ in the apparition, $g_j \geq 0$ is the amplitude and $\Delta\phi_j$ is the phase shift of the apparition. The shape $S$ is then fitted again to the aligned data of all apparitions, and the procedure is repeated three times. The amplitude $A$ is the mean of the $g_j$ weighted by the numbers of measurements, and its uncertainty is the weighted standard deviation of the $g_j$ divided by the square root of the number of apparitions. For the four test asteroids, this gives amplitudes that are all within the ranges of the published amplitudes (Appendix~\ref{app: validation}).

For an elongated body seen equator-on, the amplitude gives a lower limit of the axis ratio, $a/b \geq 10^{0.4A}$. The amplitudes in Table~\ref{tab:results} are averaged over the aspect angles of the apparitions and include phase angles up to about $30^\circ$, where the amplitudes are larger than at opposition \citep{Zappala1990}. Therefore, these limits are only indicative.

\section{Results and discussion}\label{sec: results}

We first tested both methods on four asteroids whose rotation periods are already known \citep{Warner2024}: 1951, 1963, 2134, and 2150. Their MPC brightness data were processed in the same way as those of the other asteroids, and the analysis is described in Appendix~\ref{app: validation}. The periods of both methods agree very well with the ALCDEF database (Table \ref{tab:results}). The largest differences from the ALCDEF period are found for asteroid 1951: $0.04\%$ for the S-method and $0.05\%$ for the C-method. For the other test asteroids, these differences are $0.03\%$ or smaller.

\begin{figure}
    \centering
    \includegraphics[width=1\linewidth]{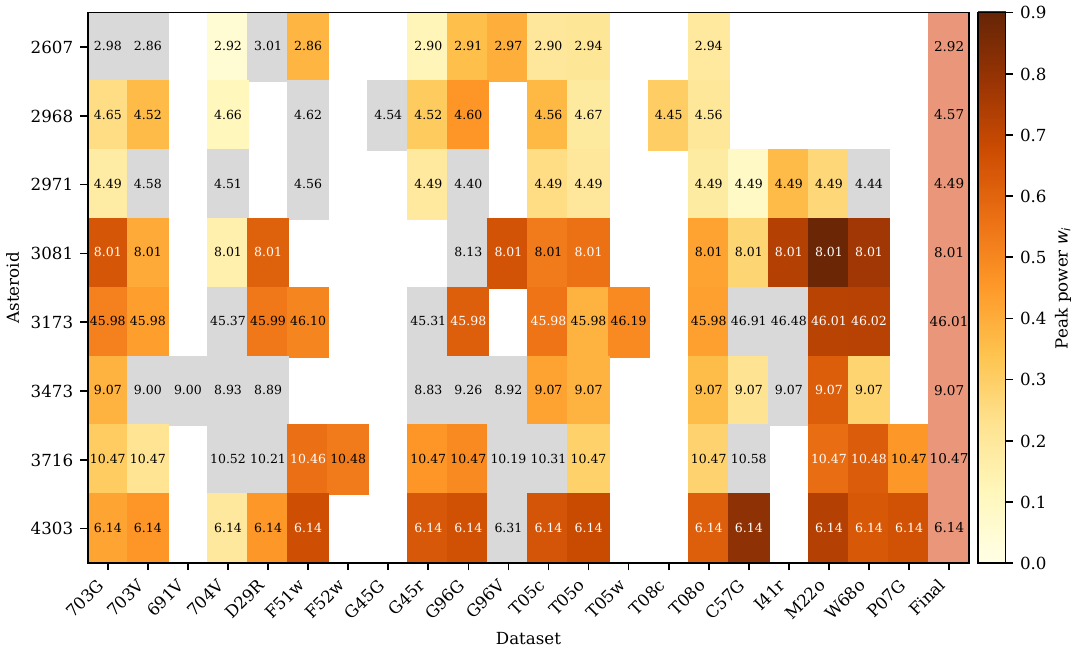}
    \caption{Summary of the S-method. Each row is an asteroid and each column a dataset. The number in a cell is the period of the dataset in hours, and the color shows its peak power $w_i$. Gray cells are datasets that do not meet the criteria of Section \ref{subsec: seperate dataset}. The last column gives the weighted mean period $P$, where the color has no physical meaning. For 2968, 2971, 3173 and 3716, which have several S-method candidates, the candidate closest to the C-method period is shown. The values of all datasets and candidates are given in the supplementary Table S1.}
    \label{fig: summary avg period}
\end{figure}

\begin{table}[ht]
\centering
\caption{Summary of the obtained results. Column~1 lists the asteroid number. Column~2 gives the rotation period reported in the ALCDEF database, where available (for 2607, see note a). Columns~3-5 give the results of the S-method: the weighted mean period $P_{\mathrm{S}}$, its mean square error $\sigma_{kv}$ (equation \ref{eq: sigma kv}) and the period significance criterion $\log\xi$ (equation \ref{eq: period significance}). Column~6 gives the period of the C-method, $P_{\mathrm{C}}$, and Column~7 its period resolution $P^2/\Delta T$, rounded up (Section~\ref{subsec: combined}). The adopted rotation period is $P_{\mathrm{C}}$, with the uncertainty of Column~7. The exception is 3173, for which no rotation period is adopted (note c). For 2968, 3173 and 3716, Columns~3-5 give the S-method candidate closest to $P_{\mathrm{C}}$. Columns~8 and~9 give the amplitude of the light curve and its uncertainty (Section~\ref{subsec: amplitude}).}
\resizebox{\textwidth}{!}{%
\begin{tabular}{|p{0.08\linewidth}|p{0.08\linewidth}|p{0.09\linewidth}|p{0.09\linewidth}|p{0.06\linewidth}|p{0.09\linewidth}|p{0.07\linewidth}|p{0.07\linewidth}|p{0.07\linewidth}|}
\hline
\textbf{Asteroid number} &
\textbf{\(P_\mathrm{ALCDEF}\) (h)} &
\textbf{\(P_\mathrm{S}\) (h)} &
\textbf{\(\sigma_{kv}\) (h)} &
\textbf{\(\log \xi\)} &
\textbf{\(P_\mathrm{C}\) (h)} &
\textbf{\(P^2/\Delta T\) (h)} &
\textbf{\(A\) (mag)} &
\textbf{\(\sigma_A\) (mag)} \\ \hline
1951 & 5.302 & 5.2999 & 0.00061 & 1.76 & 5.2995 & 0.0002 & 0.20 & 0.02 \\ \hline
1963 & 18.160 & 18.1631 & 0.0013 & 2.10 & 18.1652 & 0.0016 & 0.33 & 0.05 \\ \hline
2134 & 4.114 & 4.1144 & 0.00038 & 1.87 & 4.1149 & 0.0001 & 0.20 & 0.06 \\ \hline
2150 & 6.125 & 6.1245 & 0.00018 & 2.69 & 6.1241 & 0.0003 & 0.58 & 0.03 \\ \hline
2607 & 2.81$^{a}$ & 2.9182 & 0.0112 & 0.09 & 2.9356 & 0.0001 & 0.24 & 0.02 \\ \hline
2968 & - & 4.5725 & 0.0251 & 0.02 & 4.5596$^{b}$ & 0.0002 & 0.17 & 0.01 \\ \hline
2971 & - & 4.4902 & 0.00017 & 2.06 & 4.4908 & 0.0002 & 0.17 & 0.01 \\ \hline
3081 & - & 8.0083 & 0.00042 & 2.33 & 8.0073 & 0.0005 & 0.94 & 0.04 \\ \hline
3173 & - & 46.0137$^{c}$ & 0.0204 & 1.38 & 45.9825$^{c}$ & 0.0117 & 0.47 & 0.03 \\ \hline
3473 & - & 9.0736 & 0.00051 & 2.14 & 9.0740 & 0.0003 & 0.33 & 0.03 \\ \hline
3716 & - & 10.4727$^{d}$ & 0.0013 & 1.83 & 10.4743 & 0.0007 & 0.45 & 0.02 \\ \hline
4303 & - & 6.1367 & 0.00015 & 2.69 & 6.1363 & 0.0003 & 0.65 & 0.01 \\ \hline
\end{tabular}
}
\par\vspace{2pt}
{\footnotesize\raggedright
$^{a}$ $(2.81 \pm 0.22)$ hours from Kepler K2 photometry \citep{sergeyev2025rotation}, not listed in ALCDEF.\\
$^{b}$ Ambiguous because of daily aliasing. The S-method gives $(4.176 \pm 0.019)$ hours with $\log\xi = 0.13$ and $(3.835 \pm 0.023)$ hours with $\log\xi = -0.04$ (Section~\ref{subsec: 2968}).\\
$^{c}$ Twice the period of the strongest peak (22.99 hours) of the 0.5-240 hour search, probably a daily alias of a slower variation. The S-method also gives $(49.972 \pm 0.019)$ hours with $\log\xi = 1.44$ (Section~\ref{subsec: 3173}).\\
$^{d}$ The daily alias of $(13.4050 \pm 0.0010)$ hours gives a higher $\log\xi = 2.09$ (Section~\ref{subsec: 3716}).\par}
\label{tab:results}
\end{table}

\begin{table}[ht]
\centering
\caption{Details of the C-method. T08 was the reference observatory for all asteroids. Column~2 lists the datasets retained after step 7 of Section~\ref{sec: three} with their numbers of measurements after the outlier removal of step 2, and Column~3 gives their total. Columns~4-6 list the $HG_1G_2$ parameters of the reference fit (V band) with their formal errors (where applicable). Column~7 gives the zero-phase epoch $t_0$ of the folded light curves, and Column~8 the time span of the combined dataset.}
\scriptsize
\setlength{\tabcolsep}{2pt}
\renewcommand{\arraystretch}{0.9}
\resizebox{\textwidth}{!}{%
\begin{tabular}{|C{0.07\linewidth}|L{0.43\linewidth}|C{0.05\linewidth}|C{0.09\linewidth}|C{0.09\linewidth}|C{0.09\linewidth}|C{0.09\linewidth}|C{0.09\linewidth}|}
\hline
\textbf{Number} & \textbf{Datasets used (N)} & \textbf{N} & \textbf{H} & \textbf{$G_1$} & \textbf{$G_2$} & \textbf{$t_0$ (JD)} & \textbf{Time span} \\ \hline
1951 & 703G(209), 703V(84), C57G(1666), H45R(100), I41g(86), I41r(100), M22o(105), T05c(96), T05o(208), T08c(71), T08o(275), W68o(165) & 3165 & 14.07 & 0.09 & 0.41 & 2453767.66 & 2006.02-2024.01 \\ \hline
1963 & 689V(133), 703G(144), 703V(164), C57G(587), I41r(108), M22o(184), T05c(247), T05o(362), T05w(116), T08o(598), W68c(84), W68o(234) & 2961 & 10.45 & 0.43 & 0.00 & 2451355.92 & 1999.06-2024.01 \\ \hline
2134 & 703G(198), 703V(73), C57G(117), G45r(141), I41g(98), I41r(84), T05c(214), T05o(218), T08o(272), W68o(90) & 1505 & 13.81 & 0.00 & 1.00 & 2452751.75 & 2003.04-2024.01 \\ \hline
2150 & 703G(267), 703V(130), C57G(503), G45r(134), I41r(129), M22o(76), T05c(167), T05o(330), T08c(86), T08o(502), W68o(122) & 2446 & 12.99 & 0.00 & 0.40 & 2453494.73 & 2005.05-2024.01 \\ \hline
2607 & 703G(140), 703V(190), F51w(69), G45r(181), G96G(107), G96V(91), T05c(127), T05o(302), T08o(355) & 1562 & 13.73 $\pm$ 0.16 & 0.36 $\pm$ 0.22 & 0.27 $\pm$ 0.07 & 2452961.61 & 2003.11-2024.01 \\ \hline
2968 & 703G(102), 703V(108), F51w(73), G45r(104), G96G(71), T05c(77), T05o(130), T08c(72), T08o(282) & 1019 & 15.12 & 0.88 & 0.00 & 2453741.90 & 2006.01-2024.02 \\ \hline
2971 & 703G(258), 703V(164), F51w(81), G45r(149), G96G(96), I41r(147), M22o(105), T05c(190), T05o(258), T08o(469), W68o(205) & 2122 & 13.63 $\pm$ 0.06 & 0.47 $\pm$ 0.12 & 0.37 $\pm$ 0.06 & 2453520.67 & 2005.05-2023.12 \\ \hline
3081 & 703G(166), 703V(171), D29R(91), G96G(94), G96V(95), I41r(72), M22o(83), T05c(118), T05o(97), T08o(216), W68o(208) & 1411 & 14.32 & 0.31 & 0.69 & 2453725.76 & 2005.12-2024.01 \\ \hline
3173 & 703G(391), 703V(184), D29R(86), F51w(73), G45r(129), G96G(106), T05c(204), T05o(248), T08o(506), W68o(172) & 2099 & 13.04 $\pm$ 0.37 & 0.24 $\pm$ 0.39 & 0.34 $\pm$ 0.10 & 2452765.72 & 2003.05-2024.01 \\ \hline
3473 & 691V(71), 703G(161), 703V(190), C57G(1690), D29R(91), G45r(86), G96G(90), G96V(159), M22o(80), T05c(151), T05o(151), T08o(360), W68o(142) & 3422 & 13.76 $\pm$ 0.21 & 0.33 $\pm$ 0.37 & 0.36 $\pm$ 0.15 & 2448528.93 & 1991.09-2023.12 \\ \hline
3716 & 703G(206), 703V(213), D29R(136), F51w(71), F52w(72), G45r(113), G96G(124), G96V(117), M22o(106), P07G(72), T05c(137), T05o(355), T08o(496), W68o(81) & 2299 & 13.76 $\pm$ 0.37 & 0.53 $\pm$ 0.49 & 0.27 $\pm$ 0.15 & 2453528.68 & 2005.06-2024.06 \\ \hline
4303 & 703G(278), 703V(219), C57G(2293), D29R(120), F51w(108), G45r(124), G96G(123), G96V(71), M22o(171), P07G(159), T05c(154), T05o(286), T08o(505), W68o(191) & 4802 & 14.57 & 0.61 & 0.39 & 2453563.93 & 2005.07-2025.02 \\ \hline
\end{tabular}
}
\par\vspace{2pt}
{\footnotesize\raggedright
Datasets of Table~\ref{tab:asteroids} not used, with the criteria of step 7: 704V of 1963 and 2150, 1963 T08w, 2150 T05w and 3173 I41r (criterion 1), 704V of the other asteroids, 1951 T05w and Y00R, 2607 D29R, 2968 G45G and 3173 T05w (criterion 2 and lower power $w$ of the period), 2971, 3081 and 3716 C57G and 3173 M22o (lower $w$), and 3173 C57G and 3473 I41r (criterion 3). The results of 3173 with C57G, I41r and M22o are given in Appendix~\ref{app: 3173}.\par}
\label{tab:combined}
\end{table}

The absolute magnitudes obtained from all datasets of the C-method are given in Table~\ref{tab:absmag}. They were fitted to all DePhOCUS-corrected measurements together, without the magnitude shifts of step 5. With the DePhOCUS corrections, the offsets between the datasets are smaller than with the band offsets alone for 11 of the 12 asteroids, with rms values of $0.04$-$0.14$ mag. For 1951, 1963, 2134 and 3081, $G_1$ and $G_2$ reach the limits of the constraints, so they are not constrained by the data. The values of the eight target asteroids are compared with the literature in their subsections.

\begin{table}[ht]
\centering
\caption{Absolute magnitudes in the V band from the DePhOCUS-corrected measurements of all datasets of the C-method (Table~\ref{tab:combined}), fitted together without magnitude shifts. The uncertainties come from a bootstrap over the observing nights (one night of one dataset is one block). Column~5 gives the phase angle range of the data. Columns~6 and~7 give the rms of the median offsets of the datasets from the common fit, with the DePhOCUS corrections and with the band offsets alone.}
\begin{adjustbox}{max width=\textwidth}
\begin{tabular}{|c|c|c|c|c|c|c|}
\hline
\textbf{Asteroid} & \textbf{$H_V$} & \textbf{$G_1$} & \textbf{$G_2$} & \textbf{\shortstack{Phase\\(deg)}} & \textbf{\shortstack{rms DePhOCUS\\(mag)}} & \textbf{\shortstack{rms band\\(mag)}} \\ \hline
1951 & 14.90 $\pm$ 0.59 & 0.00 $\pm$ 0.04 & 1.00 $\pm$ 0.30 & 13.9-51.1 & 0.14 & 0.16 \\ \hline
1963 & 11.36 $\pm$ 0.06 & 0.93 $\pm$ 0.05 & 0.07 $\pm$ 0.05 & 5.6-31.2 & 0.11 & 0.16 \\ \hline
2134 & 13.50 $\pm$ 0.11 & 0.00 $\pm$ 0.02 & 0.74 $\pm$ 0.08 & 2.8-30.5 & 0.09 & 0.15 \\ \hline
2150 & 13.30 $\pm$ 0.19 & 0.20 $\pm$ 0.24 & 0.40 $\pm$ 0.09 & 1.5-33.6 & 0.09 & 0.14 \\ \hline
2607 & 13.84 $\pm$ 0.05 & 0.42 $\pm$ 0.08 & 0.29 $\pm$ 0.03 & 0.4-31.4 & 0.04 & 0.09 \\ \hline
2968 & 15.02 $\pm$ 0.12 & 0.80 $\pm$ 0.14 & 0.03 $\pm$ 0.03 & 0.9-36.6 & 0.08 & 0.12 \\ \hline
2971 & 13.53 $\pm$ 0.15 & 0.49 $\pm$ 0.15 & 0.31 $\pm$ 0.02 & 0.1-29.6 & 0.09 & 0.07 \\ \hline
3081 & 13.90 $\pm$ 0.15 & 0.00 $\pm$ 0.12 & 0.60 $\pm$ 0.08 & 0.8-29.7 & 0.08 & 0.12 \\ \hline
3173 & 12.93 $\pm$ 0.18 & 0.08 $\pm$ 0.16 & 0.38 $\pm$ 0.04 & 2.2-34.6 & 0.07 & 0.12 \\ \hline
3473 & 13.74 $\pm$ 0.07 & 0.61 $\pm$ 0.10 & 0.15 $\pm$ 0.04 & 0.4-30.7 & 0.05 & 0.12 \\ \hline
3716 & 13.62 $\pm$ 0.07 & 0.39 $\pm$ 0.10 & 0.27 $\pm$ 0.04 & 0.4-32.2 & 0.04 & 0.08 \\ \hline
4303 & 14.40 $\pm$ 0.05 & 0.13 $\pm$ 0.08 & 0.49 $\pm$ 0.04 & 0.4-31.7 & 0.06 & 0.11 \\ \hline
\end{tabular}
\end{adjustbox}
\label{tab:absmag}
\end{table}

In this section, an individual analysis of eight asteroids without a known rotation period (except for 2607, see Section \ref{Yakutia}) will be performed as well. Firstly, the properties of the asteroid will be introduced. Then the S-method will follow, where the results can already be seen in the summary plot in Fig. \ref{fig: summary avg period}. After that, the results of the C-method will be described, with supplementary power spectra and the best period light curve. The datasets, the reference phase curve parameters and the zero-phase epochs of the C-method are given in Table \ref{tab:combined}.
\subsection{(2607) Yakutia}\label{Yakutia}

2607 Yakutia, with the provisional designation 1977 NR, was first identified by N. S. Chernykh at the Crimea Observatory on 14 July 1977. Yakutia is a mid-sized X-type main belt asteroid. The diameter is 4.70 km, and the albedo value of 0.35 has been confirmed in two publications of Wide-field Infrared Survey Explorer (WISE) data enhancement NEOWISE by \citet{Mainzer2011} and \citet{Mainzer2019}. The G value adopted by \citet{Mainzer2011, Mainzer2019} is 0.15 $\pm 0.2$, while \citet{Colazo2021} determined a smaller value of 0.047 $\pm 0.188$. The published absolute magnitudes of Yakutia differ slightly. \citet{Mainzer2011, Mainzer2019} adopted the value of 13.4 mag. The lower absolute brightness is confirmed by later studies by \citet{Colazo2021}, 13.5 $\pm 0.57$, and from the Asteroid Terrestrial-impact Last Alert System (ATLAS), it is 13.62 $\pm 0.092$ \citep{Mahlke2021}. In the ALCDEF database, \citet{Warner2024} lists a diameter of 5.46 km and an albedo of 0.179, with H = 13.8 mag and G = 0.24. Our DePhOCUS-corrected data of the combined dataset give $H_V = 13.84 \pm 0.05$ (Table~\ref{tab:absmag}), close to the ALCDEF value. Recently, \citet{sergeyev2025rotation} published a rotation period of $(2.81 \pm 0.22)$ hours with an amplitude of $(0.15 \pm 0.02)$ mag from Kepler K2 photometry.

Eleven datasets with a total of 2269 brightness measurements are available for Yakutia (Table \ref{tab:asteroids}). In the S-method, eight of them meet the criteria of Section~\ref{subsec: seperate dataset}, while the 703G and 703V data give single-peaked light curves near 2.94 hours, and the D29R period is not consistent with the other datasets. The weighted mean period is $(2.918 \pm 0.011)$ hours, with $\log\xi = 0.09$, slightly above the threshold. None of the single datasets has its highest peak at this period. The highest peak powers come from the small datasets G96V, F51w and G96G (Fig. \ref{fig: summary avg period}), because the five coefficients of equation \ref{eq: fourier model} are fitted to fewer points, so the noise also gives higher peaks. For this reason, the weights of the S-method also include the number of measurements.

\begin{figure}
    \centering
    \includegraphics[width=0.5\linewidth]{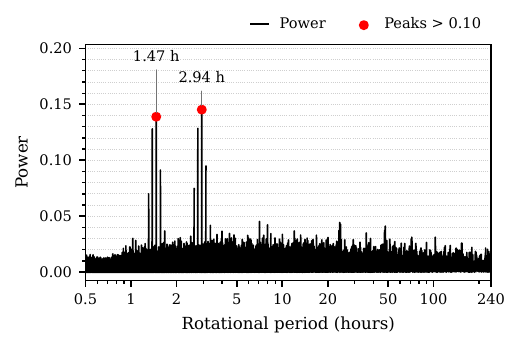}
    \caption{The power spectrum ($n=2$) of asteroid 2607 obtained from the combined brightness data over the period interval of $0.5$-$240$ hours. The red points mark the strongest peaks above the power threshold given in the legend (Section \ref{sec: L-S}), labeled with their periods. Weaker peaks close to a marked peak are not marked.}
    \label{fig: 2607 power spectra}
\end{figure}

\begin{figure}
    \centering
    \includegraphics[width=0.5\linewidth]{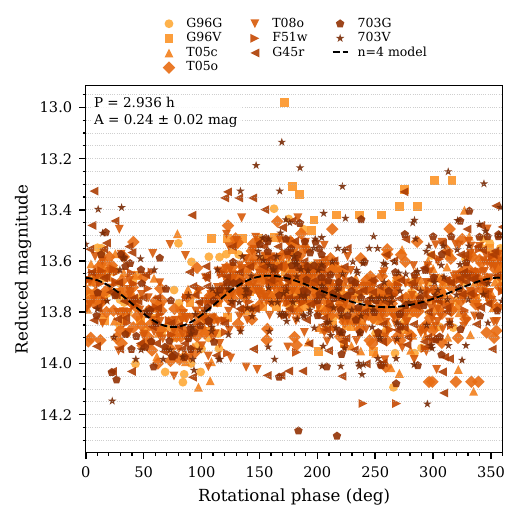}
    \caption{The light curve of asteroid 2607 obtained from the combined brightness data, folded with the period of $2.936$ hours (phase 0 at JD 2452961.61). Symbols distinguish the datasets. The dashed line is the $n=4$ Fourier model. The amplitude $A$ is the mean of the amplitudes of the individual apparitions (Section~\ref{subsec: amplitude}), so it can be larger than the peak-to-peak range of the model of the combined data. The same conventions are used in the light curve figures of the other asteroids.}
    \label{fig: 2607 light curve}
\end{figure}

As for the C-method, the reference Mauna Loa observatory with the $HG_1G_2$ model resulted in best fit values of $H = 13.73 \pm 0.16$, $G_1 = 0.36 \pm 0.22$ and $G_2 = 0.27 \pm 0.07$ (Table \ref{tab:combined}). Two dominant peaks were obtained from the Lomb-Scargle method for the 1562 combined measurements, seen in Fig. \ref{fig: 2607 power spectra}: $2.936$ and $1.468$ hours, with practically the same power ($w = 0.15$ and $0.14$). The weaker peaks next to them, at about $2.77$ and $3.13$ hours and at half of these values ($w \leq 0.13$), are their daily aliases ($48/2.77 - 48/2.936 \approx 48/2.936 - 48/3.13 \approx 1$ days$^{-1}$). The $1.47$ hour light curve has only one maximum and one minimum, so this period was discarded. Further on in the analysis, there will always be a period two times smaller than the real one, as the light curve can be folded in half with minimal accuracy reduction. This effect will not be explicitly commented on in the further sections. It can be confidently stated that the period is $2.936$ hours shown in Fig. \ref{fig: 2607 light curve}.

The rotation periods from both methods, $(2.918 \pm 0.011)$ hours and $2.936$ hours, agree with each other within twice the error of the S-method, and with the period of \citet{sergeyev2025rotation} within its error. The amplitude of $(0.24 \pm 0.02)$ mag is larger than the K2 value, which was measured in a single apparition with a different aspect. Due to the low significance criterion of the S-method ($\log\xi = 0.09$, close to the threshold), the C-method period of $(2.9356 \pm 0.0001)$ hours is adopted.

\subsection{(2968) Iliya} \label{subsec: 2968}
The S*-type asteroid Iliya, provisional designation 1978 QJ, was discovered by N. S. Chernykh at the Crimea Observatory on 31 August 1978. \citet{Nugent2015} found a diameter of $(4.46 \pm 0.96)$ km and an albedo of $0.13 \pm 0.08$, and \citet{Mainzer2016} confirmed these values. Greatly different results, $(6.24 \pm 0.92)$ km for the diameter and $0.055 \pm 0.011$ for the albedo, were obtained by \citet{AliLagoa2017}. There are different reports for the absolute magnitude: 14.30 $\pm 0.92$ is given by \citet{Carry2016}, 14.8 by \citet{AliLagoa2017}, while \citet{Mainzer2016} adopted a value of 14.6 mag from the MPC using $G = 0.15$. With a higher accuracy, \citet{Mahlke2021} calculated the H value of 14.44 $\pm 0.104$ using the $HG_1G_2$ method. \citet{Warner2024} gives a diameter of 6.21 km and an albedo of 0.055 in the ALCDEF database, with H = 14.8 mag and G = 0.15. Our DePhOCUS-corrected data give $H_V = 15.02 \pm 0.12$ (Table~\ref{tab:absmag}), fainter than these values.

For Iliya, 1431 brightness measurements from eleven datasets were analyzed (Table \ref{tab:asteroids}). The largest dataset, T08o with 297 measurements, has its highest peak at 4.56 hours ($w = 0.20$). The S-method gives three candidates, which are daily aliases of each other: $(4.572 \pm 0.025)$, $(4.176 \pm 0.019)$ and $(3.835 \pm 0.023)$ hours, with $\log\xi$ of $0.02$, $0.13$ and $-0.04$. All three are close to the threshold, so the S-method alone cannot decide between them.

\begin{figure}
    \centering
    \includegraphics[width=0.5\linewidth]{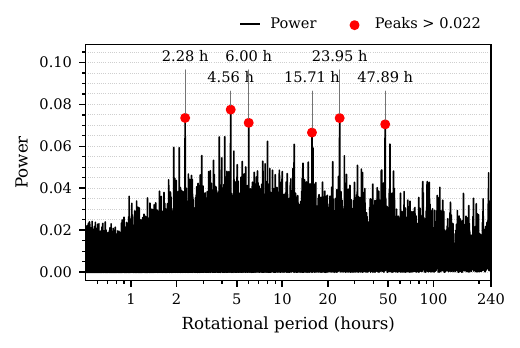}
    \caption{The power spectrum ($n=2$) of asteroid 2968 obtained from the combined brightness data, with the same conventions as in Fig. \ref{fig: 2607 power spectra}.}
    \label{fig: 2968 power spectra}
\end{figure}

\begin{figure}
    \centering
    \includegraphics[width=0.5\linewidth]{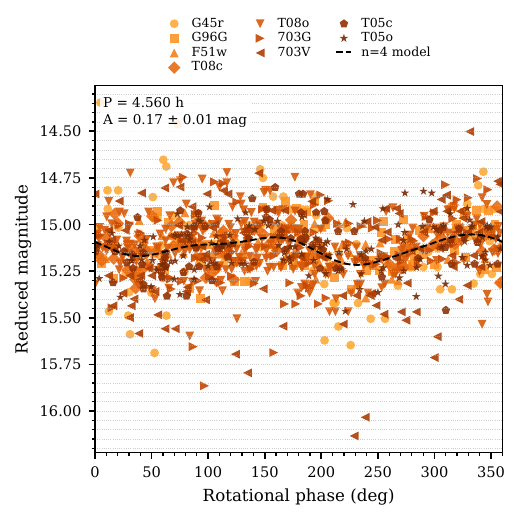}
    \caption{The light curve of asteroid 2968 obtained from the combined brightness data, folded with the period of $4.560$ hours (phase 0 at JD 2453741.90).}
    \label{fig: 2968 light curve}
\end{figure}

The C-method analysis of 1019 measurements gave the strongest peak of the power spectrum at $4.560$ hours ($w = 0.078$, Fig. \ref{fig: 2968 power spectra}). Its daily aliases at $3.831$ and $4.163$ hours ($48/3.831 - 48/4.163 \approx 48/4.163 - 48/4.560 \approx 1$ days$^{-1}$) are only slightly weaker ($w = 0.065$). The power spectrum also has peaks at about $6$, $24$ and $48$ hours ($w \approx 0.07$), which are the common values of a window function for combined datasets, and weaker peaks at $15.7$ and $51.5$ hours ($w \leq 0.067$). The light curve folded with $4.560$ hours has two maxima and two minima (Fig. \ref{fig: 2968 light curve}), with an amplitude of $(0.17 \pm 0.01)$ mag. However, the power of this period is low and close to that of its aliases and of the window function peaks, so the alias periods of $3.83$ and $4.16$ hours cannot be excluded with the present data, and the period of 2968 remains ambiguous.

\subsection{(2971) Mohr}
The S-type asteroid Mohr (provisional designation: 1980 YL) was discovered by A. Mrkos at the Klet Observatory on 30 December 1980. \citet{Mainzer2011} derived a diameter of $(5.654 \pm 0.024)$ km and an albedo of $0.2211 \pm 0.0239$ from enhanced Wide-field Infrared Survey Explorer (WISE) data. Similar results were obtained by \citet{Mainzer2019}: diameter of $(5.167 \pm 0.056)$ km and albedo of 0.263 $\pm 0.033$. \citet{Mainzer2011} and \citet{Mainzer2019} adopted the absolute magnitude $13.5 \pm 0.1$. \citet{Veres2015} obtained a slightly larger absolute magnitude of $13.68 \pm 0.29$. However, \citet{Mahlke2021} obtained $12.95 \pm 0.044$ using the $HG_1G_2$ method, unlike previous authors who used the HG method in the G=0.15 approximation. The ALCDEF database gives H = 13.39 mag and G = 0.24 for Mohr, with a diameter of 5.09 km and an albedo of 0.3. Our DePhOCUS-corrected data give $H_V = 13.53 \pm 0.15$ (Table~\ref{tab:absmag}), in agreement with \citet{Mainzer2011} and \citet{Veres2015}.

\begin{figure}
    \centering
    \includegraphics[width=0.5\linewidth]{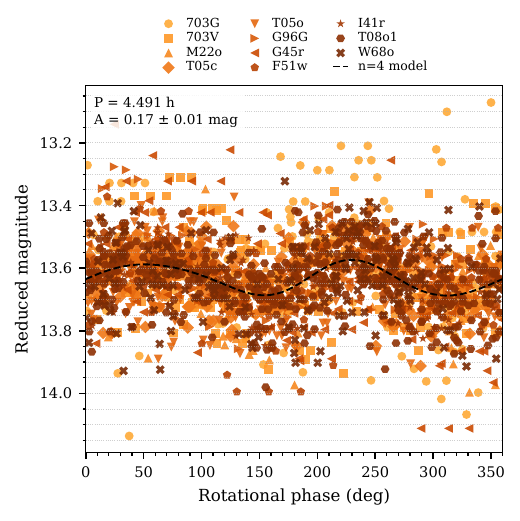}
    \caption{The light curve of asteroid 2971 obtained from the combined brightness data, folded with the period of $4.491$ hours (phase 0 at JD 2453520.67).}
    \label{fig: 2971 light curve}
\end{figure}

Mohr has 4740 measurements in thirteen datasets (Table \ref{tab:asteroids}). The power spectra of the largest datasets show two maxima, at about 4.5 and 4.8 hours. In the S-method, eight datasets give periods between $4.4897$ and $4.4908$ hours, and their weighted mean is $(4.4902 \pm 0.0002)$ hours, with $\log\xi = 2.06$. The other candidate, near 4.8 hours, gives only $\log\xi = 0.10$. The mean peak power of Mohr, $W_{av} = 0.21$, is the lowest of the sample. This is partly because Mohr has large datasets, and the noise peaks are lower when more measurements are fitted (Section~\ref{Yakutia}). For example, the TESS dataset, with almost 2000 measurements, gives only $w = 0.085$. It is also due to the small amplitude of the asteroid, seen in Fig. \ref{fig: 2971 light curve}. The Lomb-Scargle method is based on the $\chi^2$ minimization formalism, and asteroids with intrinsically low rotational light curve amplitudes produce a smaller reduction in $\chi^2$ relative to a constant model, resulting in lower power. Nevertheless, the periods of the individual datasets agree well, which gives a small $\sigma_{kv}$ and a high $\xi$.

\begin{figure}
    \centering
    \includegraphics[width=0.5\linewidth]{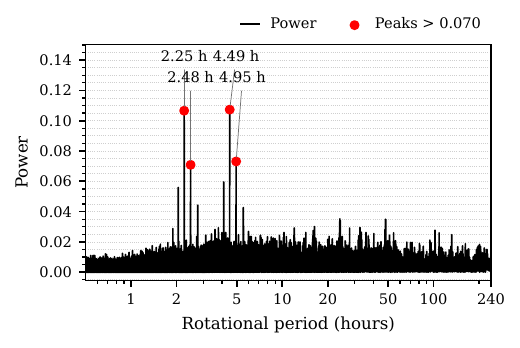}
    \caption{The power spectrum ($n=2$) of asteroid 2971 obtained from the combined brightness data, with the same conventions as in Fig. \ref{fig: 2607 power spectra}.}
    \label{fig: 2971 power spectra}
\end{figure}

The period was confirmed by the C-method. The TESS dataset (C57G) was not used in the C-method, because it lowers the power of the period from $0.11$ to $0.08$ and shifts the period by $0.0008$ hours (step 7 in Section~\ref{sec: three}). The C-method combines 2122 measurements of eleven datasets (Table \ref{tab:combined}). Only one viable period of $4.491$ hours was found in Fig. \ref{fig: 2971 power spectra}. Its daily alias at $4.954$ hours ($48/4.491 - 48/4.954 \approx 1$ days$^{-1}$) has a lower power ($w = 0.07$) than the period ($w = 0.11$). The light curve is visualized in Fig. \ref{fig: 2971 light curve}, and its amplitude is $(0.17 \pm 0.01)$ mag.

\subsection{(3081) Martinuboh} \label{subsec: 3081}

The S*-type asteroid Martinuboh (provisional designation: 1971 UP) was discovered by L. Kohoutek at the Bergedorf Observatory on 26 October 1971. In detail, the asteroid has been studied only by \citet{Mahlke2021}. ATLAS dual band photometry data analysis with $HG_1G_2$ method gives two absolute magnitude values: 14.23 $\pm 0.142$ from cyan band measurements and 14.07 $\pm 0.024$ from orange band. A diameter of 4.86 km is mentioned only in the ALCDEF database by \citet{Warner2024}. The database also gives H = 13.93, G = 0.24 and albedo 0.20. Our DePhOCUS-corrected data give $H_V = 13.90 \pm 0.15$ (Table~\ref{tab:absmag}), close to the ALCDEF value, although $G_1$ reaches the limit of the constraints.

Martinuboh has 2664 measurements in thirteen datasets (Table \ref{tab:asteroids}). Our first data files contained only a reduced set of these data (phase angles above $7^\circ$, without TESS), so all data of this asteroid were taken from the MPC observation records (Section~\ref{sec: Data processing}). In the S-method, twelve of the thirteen datasets give periods between $8.0071$ and $8.0110$ hours, and their weighted mean is $(8.0083 \pm 0.0004)$ hours, with $\log\xi = 2.33$. The G96G data give $8.131$ hours, which is not consistent with the other datasets.

\begin{figure}
    \centering
    \includegraphics[width=0.5\linewidth]{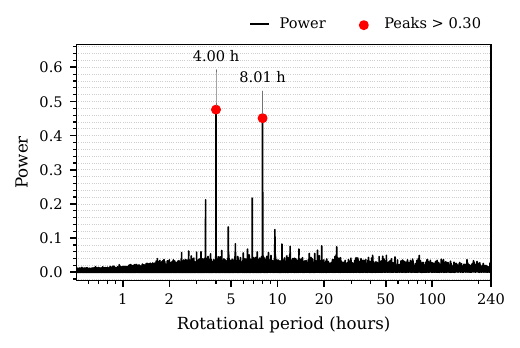}
    \caption{The power spectrum ($n=2$) of asteroid 3081 obtained from the combined brightness data, with the same conventions as in Fig. \ref{fig: 2607 power spectra}.}
    \label{fig: 3081 power spectra}
\end{figure}

\begin{figure}
    \centering
    \includegraphics[width=0.5\linewidth]{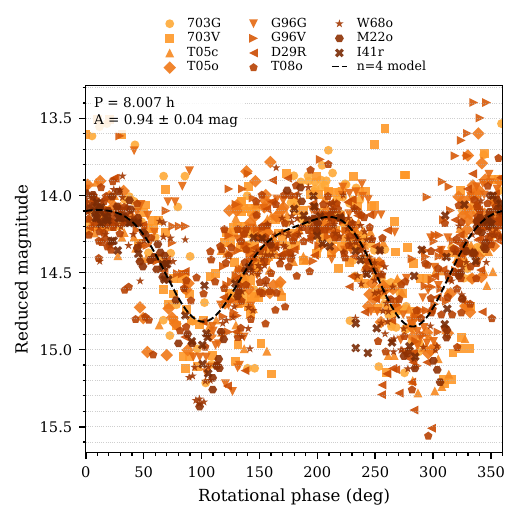}
    \caption{The light curve of asteroid 3081 obtained from the combined brightness data, folded with the period of $8.007$ hours (phase 0 at JD 2453725.76).}
    \label{fig: 3081 light curve}
\end{figure}

The obtained period is supported by the C-method, which uses 1411 measurements of eleven datasets (Table \ref{tab:combined}). The TESS data, all at phase angles below $7^\circ$, lowered the power of the period from $0.44$ to $0.36$ in the test of step 7 and were not used. The C-method also produced only one viable period of $8.007$ hours ($w = 0.45$) and its half ($w = 0.48$), seen in Fig. \ref{fig: 3081 power spectra}. It must be noted that this period is close to a common window function peak at $8$ hours. However, the peak is clearly separated from it, as the power at exactly $8.000$ hours is close to zero ($w = 0.02$). The peaks at $24$ and $12$ hours ($w \approx 0.07$) belong to the window function. The weaker peaks at $6.86$ and $3.43$ hours ($w \le 0.22$) are the daily aliases of $8.007$ and $4.004$ hours ($48/6.860 - 48/8.007 \approx 1$ days$^{-1}$). Due to the fact that these peaks have less than half of the power of the $8.007$ hour period, and it has a strong counterpart at the half period of $4.004$ hours, the $8.007$ hour period can be safely assumed to be the real period. It can be further justified by visual inspection in Fig. \ref{fig: 3081 light curve}, where it has a distinct form with two maxima and two minima, which is not the case for the alias period. The light curve has an amplitude of $(0.94 \pm 0.04)$ mag, which gives a lower limit of the axis ratio of $a/b \geq 2.4$.

\subsection{(3173) McNaught} \label{subsec: 3173}
The S*-type asteroid 3173 (provisional designation: 1981 WY) was first identified by E. Bowell at Lowell Observatory's Anderson Mesa Station on 24 November 1981. Wide-field Infrared Survey Explorer (WISE) data give a diameter of $(6.379 \pm 0.025)$ km and an albedo of $0.2283 \pm 0.0649$ \citep{Mainzer2011}. In 2019, \citet{Mainzer2019} corrected the diameter and albedo to $(6.026 \pm 0.053)$ km and $0.286 \pm 0.024$. The G value in both Mainzer papers was fixed to 0.15, and the H value was adopted from the MPC. The absolute magnitude values in papers from 2011 till 2021 change little, from $13.2 \pm 0.1$ \citep{Mainzer2011, Mainzer2019} to $13.26 \pm 0.47$ \citep{Veres2015} and $13.24 \pm 0.23$ \citep{Mahlke2021}. However, ATLAS dual band photometry data analysis with $HG_1G_2$ method from orange band photometry gives a brighter value: H = $12.67 \pm 0.145$ \citep{Mahlke2021}. \citet{Warner2024} adopts H = 13.18 mag and G = 0.24 in the ALCDEF database, with a diameter of 6.87 km and an albedo of 0.20. Our DePhOCUS-corrected data give $H_V = 12.93 \pm 0.18$ (Table~\ref{tab:absmag}), between these values.

McNaught has 5956 measurements in fifteen datasets (Table \ref{tab:asteroids}). The S-method gives two candidates. Eleven datasets give $(46.014 \pm 0.020)$ hours with $\log\xi = 1.38$, and twelve datasets give $(49.972 \pm 0.019)$ hours with $\log\xi = 1.44$. The TESS dataset gives neither of them, and its strongest peaks are close to the upper end of the search range, near 236 hours.

\begin{figure}
    \centering
    \includegraphics[width=0.5\linewidth]{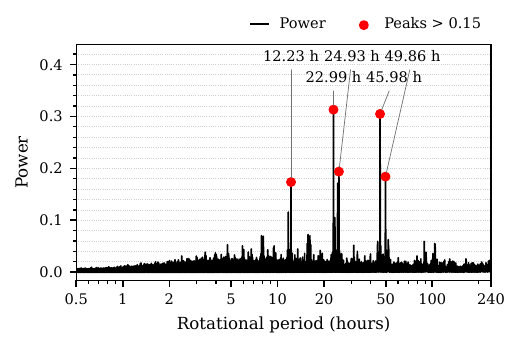}
    \caption{The power spectrum ($n=2$) of asteroid 3173 obtained from the combined brightness data, with the same conventions as in Fig. \ref{fig: 2607 power spectra}.}
    \label{fig: 3173 power spectra}
\end{figure}

\begin{figure}
    \centering
    \includegraphics[width=0.5\linewidth]{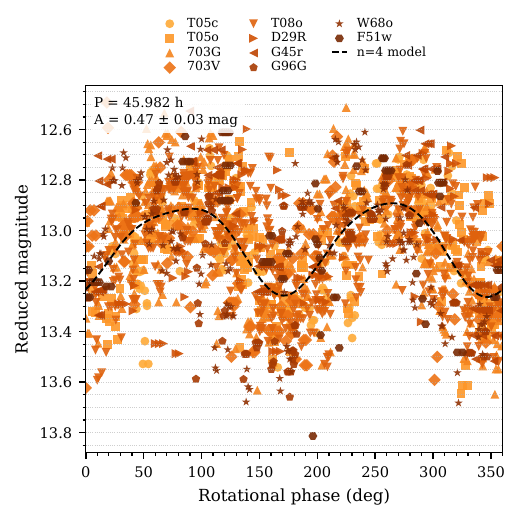}
    \caption{The light curve of asteroid 3173 obtained from the combined brightness data, folded with the period of $45.982$ hours, twice the period of the strongest peak at $22.991$ hours (phase 0 at JD 2452765.72).}
    \label{fig: 3173 light curve}
\end{figure}

The TESS (C57G), I41r, M22o and T05w datasets and the LINEAR data were not used in the C-method (Table \ref{tab:combined}), which uses 2099 measurements of ten datasets. The C-method gives several potential periods (Fig. \ref{fig: 3173 power spectra}). This is expected, as for long-period asteroids, the temporal coverage is naturally less populated. The $45.98$ hour period (Fig. \ref{fig: 3173 light curve}) has a higher power ($w = 0.30$) than the other candidates ($w \leq 0.19$), apart from its half period of $22.99$ hours ($w = 0.31$). However, the peak at $49.9$ hours is a daily alias of the $45.98$ hour period, as $48/45.98 + 48/49.9 \approx 2$ days$^{-1}$. The $49.9$ hour period is favored by four of the five parts of the data around the oppositions with at least 50 measurements, when they are analyzed separately. A C-method search extended to $1500$ hours shows stronger peaks at $572$ hours ($w = 0.43$) and $1143$ hours ($w = 0.42$). The short periods are daily aliases of this slow variation, as $48/45.98 - 48/1143 \approx 1$ days$^{-1}$ (Appendix~\ref{app: 3173}). The TESS measurements of 2020 cover two stretches of 9 and 12 days without daily gaps and show the slow variation, but no $46$ or $50$ hour variation larger than about $0.05$ mag. When the TESS data are added to the combined data, the $46$ and $50$ hour peaks almost disappear ($w \leq 0.05$). Therefore, the $45.98$ hour period in Table \ref{tab:results} is only the strongest candidate of the 0.5-240 hour search, and it is probably not the rotation period. The rotation period of 3173 remains undetermined and may be much longer.

\subsection{(3473) Sapporo}

The S-type asteroid 3473 (provisional designation: A924 EG) was discovered by K. Reinmuth at the Heidelberg Observatory on 7 March 1924. Its diameter, $(4.28 \pm 0.75)$ km, and albedo, $0.29 \pm 0.13$, have been measured only by \citet{Nugent2016}. They used data of the second year of the WISE/NEOWISE mission and took the H value from the MPC. In literature, the absolute magnitude changes by half a magnitude from $14.14 \pm 0.59$ \citep{Veres2015}, through 13.90 \citep{Nugent2016} to $13.87 \pm 0.169$ and $13.61 \pm 0.040$ \citep{Mahlke2021} from ATLAS cyan and orange bands, respectively. In the ALCDEF database \citep{Warner2009, Warner2024}, the diameter is 4.68 km and the albedo 0.26, with H = 13.73 and G = 0.24. Our DePhOCUS-corrected data give $H_V = 13.74 \pm 0.07$ (Table~\ref{tab:absmag}), close to the ALCDEF value.

Sapporo has 4459 measurements in fifteen datasets (Table \ref{tab:asteroids}). The largest dataset, the 1817 TESS measurements reported in the ``G'' band, has its strongest peak at 9.074 hours ($w = 0.22$). In the S-method, seven datasets meet the criteria, and their weighted mean period is $(9.0736 \pm 0.0005)$ hours, with $\log\xi = 2.14$. The other datasets give a single-peaked light curve (691V), peaks below the $5\sigma$ level (D29R and I41r), or periods that are not consistent with the other datasets (criterion 3, 703V, 704V, G45r, G96G and G96V).

\begin{figure}
    \centering
    \includegraphics[width=0.5\linewidth]{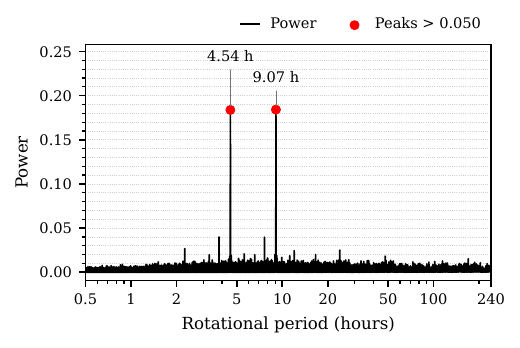}
    \caption{The power spectrum ($n=2$) of asteroid 3473 obtained from the combined brightness data, with the same conventions as in Fig. \ref{fig: 2607 power spectra}.}
    \label{fig: 3473 power spectra}
\end{figure}

\begin{figure}
    \centering
    \includegraphics[width=0.5\linewidth]{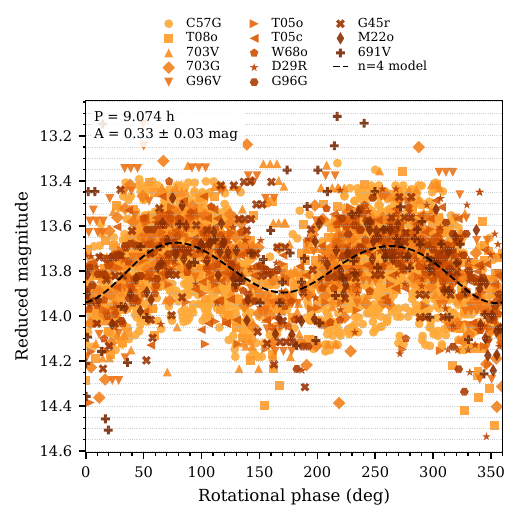}
    \caption{The light curve of asteroid 3473 obtained from the combined brightness data, folded with the period of $9.074$ hours (phase 0 at JD 2448528.93).}
    \label{fig: 3473 light curve}
\end{figure}

This agrees excellently with the C-method, as there is only one viable peak in Fig. \ref{fig: 3473 power spectra}, which matches the period of $9.074$ hours ($w = 0.18$). The daily alias at $7.63$ hours reaches only $w = 0.04$. The combined dataset consists of 3422 measurements of thirteen datasets (Table \ref{tab:combined}). The TESS data were retained, as their phase curve is consistent with that of the reference observatory. The I41r dataset was discarded by criterion (3) of step 7. In one night of March 2018, its 91 measurements change smoothly by $0.7$ mag within one hour, more than three times the largest change the light curve allows in this time, and they form a separate branch below the other datasets. The light curve can be seen in Fig. \ref{fig: 3473 light curve}, and its amplitude is $(0.33 \pm 0.03)$ mag.

\subsection{(3716) Petzval} \label{subsec: 3716}

The asteroid 3716 (provisional designation: 1980 TG) was discovered by A. Mrkos at the Klet Observatory on 2 October 1980. The asteroid is a Nysa-Polana complex object with an unclear taxonomic class. The diameter of the asteroid is $(4.691 \pm 0.533)$ km and albedo is $0.35 \pm 0.152$, which have been determined by \citet{Mainzer2016} from the NEOWISE-Reactivation Year 1 operations data, with the fixed parameter G = 0.15. A brighter absolute magnitude of $13.15 \pm 0.40$ is given by \citet{Veres2015}. Later investigations give fainter H values: 13.40 was adopted by \citet{Mainzer2016}, and $13.32 \pm 0.033$ was obtained by \citet{Mahlke2021}. The ALCDEF database lists a diameter of 4.57 km and an albedo of 0.28 for Petzval, with H = 13.70 mag and G = 0.24. Our DePhOCUS-corrected data give $H_V = 13.62 \pm 0.07$ (Table~\ref{tab:absmag}), close to the ALCDEF value.

Petzval has 3123 measurements in sixteen datasets (Table \ref{tab:asteroids}). Their power spectra show notable peaks at about 30, 18.6, 13.4 and 10.5 hours, and the light curves folded with these periods have two maxima and two minima. To select the correct rotation period, the period significance criterion is used (equation \ref{eq: period significance}). The S-method gives $(30.389 \pm 0.026)$, $(18.617 \pm 0.010)$, $(13.4050 \pm 0.0010)$ and $(10.4727 \pm 0.0013)$ hours, with $\log\xi$ of $0.89$, $1.17$, $2.09$ and $1.83$. The two strongest candidates, $13.405$ and $10.473$ hours, are daily aliases of each other ($48/10.473 - 48/13.405 \approx 1$ days$^{-1}$). The $13.405$ hour period has the higher $\log\xi$, but both values are well above the threshold, so the S-method alone cannot decide between them.

\begin{figure}
    \centering
    \includegraphics[width=0.5\linewidth]{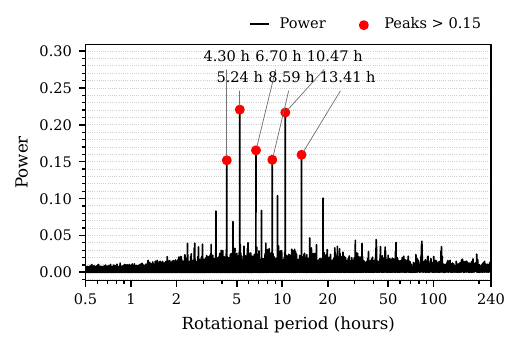}
    \caption{The power spectrum ($n=2$) of asteroid 3716 obtained from the combined brightness data, with the same conventions as in Fig. \ref{fig: 2607 power spectra}.}
    \label{fig: 3716 power spectra}
\end{figure}

\begin{figure}
    \centering
    \includegraphics[width=0.5\linewidth]{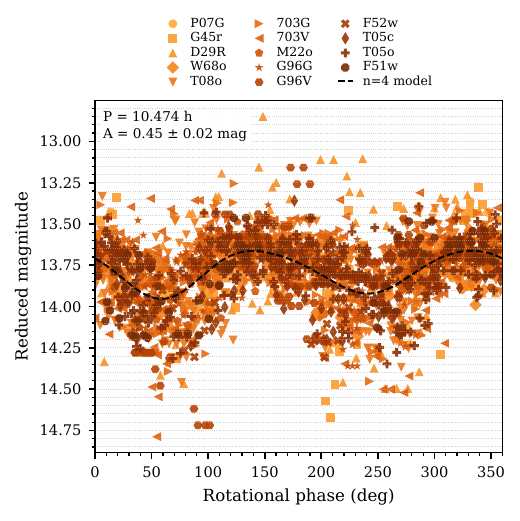}
    \caption{The light curve of asteroid 3716 obtained from the combined brightness data, folded with the period of $10.474$ hours (phase 0 at JD 2453528.68).}
    \label{fig: 3716 light curve}
\end{figure}

The C-method analysis of 2299 measurements from fourteen datasets faced similar difficulties, obtaining several peaks in the power spectrum seen in Fig. \ref{fig: 3716 power spectra}. Nevertheless, the highest peaks correspond to the $10.474$ hour period and its half ($w = 0.22$), while the $13.405$ hour alias and its half reach only $w = 0.16$-$0.17$. The peaks at $8.59$ and $4.30$ hours ($w = 0.15$) are the daily aliases on the other side ($48/8.594 - 48/10.474 \approx 1$ days$^{-1}$). The light curve folded with $10.474$ hours is seen in Fig. \ref{fig: 3716 light curve}, and its amplitude is $(0.45 \pm 0.02)$ mag. Because the C-method clearly prefers the $10.474$ hour period, it is the most likely one, but the alias at $13.405$ hours, which has the higher $\log\xi$ in the S-method, cannot be excluded completely. The strong presence of the 13 hour period is explained by the sampling, as it is the daily alias of the $10.474$ hour period. Therefore, the 13 hour period does not by itself indicate that the asteroid is tumbling, although this cannot be excluded without further analysis.

\subsection{(4303) Savitskij} \label{subsec: 4303}

The asteroid 4303 (provisional designation: 1973 SZ3) was discovered by L. V. Zhuravleva at the Crimea Observatory on 25 September 1973. It is an S*-type asteroid. The diameter of the asteroid is $(4.145 \pm 0.145)$ km, and the albedo is $0.236 \pm 0.025$, which were obtained by \citet{Mainzer2011, Mainzer2019}. The absolute magnitude adopted in these works was $14.1 \pm 0.1$ using the HG method, and $H = 14.07 \pm 0.089$ was obtained by \citet{Mahlke2021}, using the $HG_1G_2$ method with orange band ATLAS brightness. A very different H value of $14.40 \pm 0.071$ was obtained by \citet{Mahlke2021} from analysis of cyan band ATLAS data. The diameter of 4.08 km given by \citet{Warner2024} in the ALCDEF database is slightly smaller. The database also gives an albedo of 0.20, H = 14.31 mag and G = 0.24. Our DePhOCUS-corrected data give $H_V = 14.40 \pm 0.05$ (Table~\ref{tab:absmag}), close to the cyan band value of \citet{Mahlke2021} and to the ALCDEF value.

Savitskij has 5434 measurements in fifteen datasets (Table \ref{tab:asteroids}), including 2352 TESS measurements in the ``G'' band. In the S-method, fourteen datasets give periods between $6.1362$ and $6.1377$ hours, and their weighted mean is $(6.1367 \pm 0.0001)$ hours, with $\log\xi = 2.69$. Only the G96V data, with 76 measurements, give a different peak. Because of the large amplitude, every dataset gives a strong peak, and the mean peak power is high, $W_{av} = 0.59$ (Fig. \ref{fig: summary avg period}).

\begin{figure}
    \centering
    \includegraphics[width=0.5\linewidth]{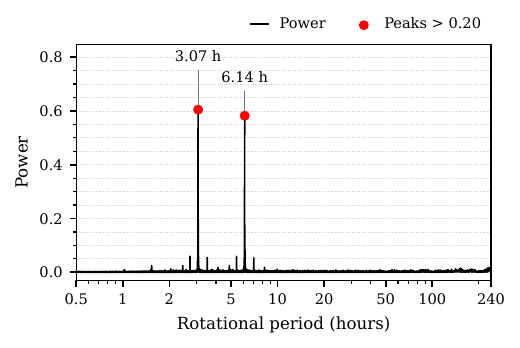}
    \caption{The power spectrum ($n=2$) of asteroid 4303 obtained from the combined brightness data, with the same conventions as in Fig. \ref{fig: 2607 power spectra}.}
    \label{fig: 4303 power}
\end{figure}

\begin{figure}
    \centering
    \includegraphics[width=0.5\linewidth]{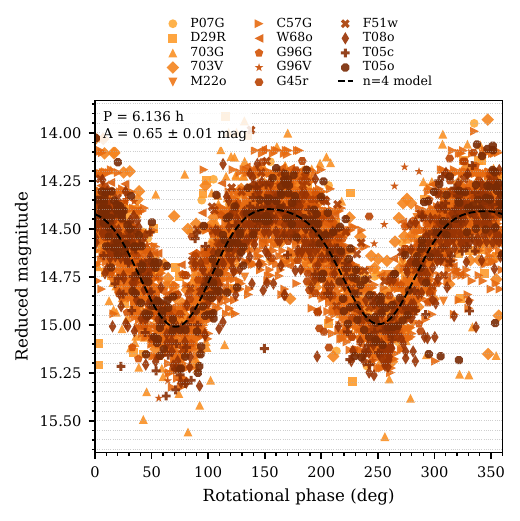}
    \caption{The light curve of asteroid 4303 obtained from the combined brightness data, folded with the period of $6.136$ hours (phase 0 at JD 2453563.93).}
    \label{fig: 4303 lc}
\end{figure}

From the C-method analysis of 4802 measurements from fourteen datasets, including TESS, a clear period is obtained, seen in Fig. \ref{fig: 4303 power} with a period of $6.136$ hours. The daily aliases of this period are much weaker, with $w \le 0.06$. The C-method period agrees with the S-method period within $0.0005$ hours. When the eight oppositions with at least 50 measurements are analyzed separately, seven of them give periods between $6.1361$ and $6.1371$ hours. Only the poorly sampled opposition of 2013 (82 measurements) gives a different period. The light curve of the 4303 asteroid can be seen in Fig. \ref{fig: 4303 lc}, and its amplitude is $(0.65 \pm 0.01)$ mag, which gives $a/b \geq 1.8$.

\section{Summary}\label{sec: summary}

We analyzed the rotation periods of eight main belt asteroids with an Earth MOID below 1 AU, using heterogeneous photometry reported to the Minor Planet Center. The periods of six asteroids are new, the period of (2607) Yakutia agrees with a recently published value \citep{sergeyev2025rotation} and is more precise, and the rotation period of (3173) McNaught remains undetermined. Two complementary methods were applied. The separate dataset analysis (S-method) builds on our earlier work \citep{Eglitis2023} and adds the period significance criterion. The combined dataset analysis (C-method) is a new data combination pipeline based on the $HG_1G_2$ photometric model and the DePhOCUS bias corrections. These methods enabled the effective use of brightness measurements from multiple observatories and bands to construct consistent light curves and obtain new rotation periods. Both methods were tested on four asteroids with known rotation periods and reproduced the ALCDEF periods within 0.05\%.

Together with the systematic photometric corrections and statistical filtering of Sections~\ref{sec: Data processing} and \ref{sec: three}, the Lomb-Scargle method reliably finds rotation periods in sparse data with uneven sampling over long time baselines. In this work, the combined datasets contained 1019-4802 measurements (Table~\ref{tab:combined}). Both the S-method and the C-method produced consistent results. The C-method notably reduced noise and aliasing, improving period accuracy and enabling the identification of true periodicities, even with observations that have large uncertainties. The S-method proved useful for comparing results, while the metric defined in equation~\ref{eq: period significance}, the period significance criterion, is particularly valuable when multiple possible periods are obtained from either method. The DePhOCUS corrections also give absolute magnitudes in the V band, and they reduce the offsets between the datasets.

The C-method also proved to be a reliable approach and offers clear advantages, including a higher signal-to-noise ratio, reduced aliasing, more accurate final periods, and straightforward automation. For large datasets of limited precision and uneven time sampling, the best strategy is therefore to build a common phase diagram from all data first. The result should then be checked by analyzing the individual observatories separately. Certainly, the S-method offers other advantages, such as the possibility to report indicative errors and apply the period significance criterion.

Offsets between the datasets that change with time, phase angle or color are not removed by the magnitude shifts. They remain in the combined light curves, where they add scatter and can add spurious power at long periods. Due to the heterogeneous accuracy of the brightness data being analyzed, it is not useful to rely on either method alone. Rather, both methods complement each other and should be used simultaneously to determine the true rotation period. This is especially useful in cases where an asteroid has multiple possible rotation period values that fulfill the criteria of Section~\ref{subsec: seperate dataset}. For asteroid 2968, the daily aliases at 3.83 and 4.16 hours remain possible alternatives to the adopted period of 4.560 hours, and for 3716 the daily alias at 13.405 hours cannot be excluded completely. For 3173, the 45.98 and 49.9 hour peaks are probably daily aliases of a slower brightness variation of about 572 or 1143 hours, and the rotation period is not determined. Both methods were tested on main belt asteroids only. For NEOs, the phase angle and viewing geometry change quickly, so the methods are not directly applicable. Their application to NEOs requires a separate study.

The next steps are to automate the C-method pipeline so it can be applied to a large sample of asteroids, and to extend the calibration with color and phase angle dependent terms, building on the DePhOCUS corrections applied here.

\printcredits

\section*{Acknowledgements}
\label{Acknowledgements}
This research is funded by the European Regional Development Fund, project ``ERA Chair in Astrophysics, Instrumentation, Ground Segment Technologies and Space Photonics at the University of Latvia'', project No. 1.1.1.5/2/24/A/004.

\section*{Declaration of generative AI and AI-assisted technologies in the manuscript preparation process}
During the preparation of this work the authors used large language models (LLM) in order to edit the text for language and clarity and to assist with writing the analysis code. After using this tool, the authors reviewed and edited the content as needed and take full responsibility for the content of the published article.

\appendix
\makeatletter
\renewcommand\@seccntformat[1]{Appendix~\csname the#1\endcsname\@seccntDot\hskip 0.5em}
\makeatother

\section{Validation with asteroids of known period}\label{app: validation}

\begin{figure}
    \centering
    \includegraphics[width=0.9\linewidth]{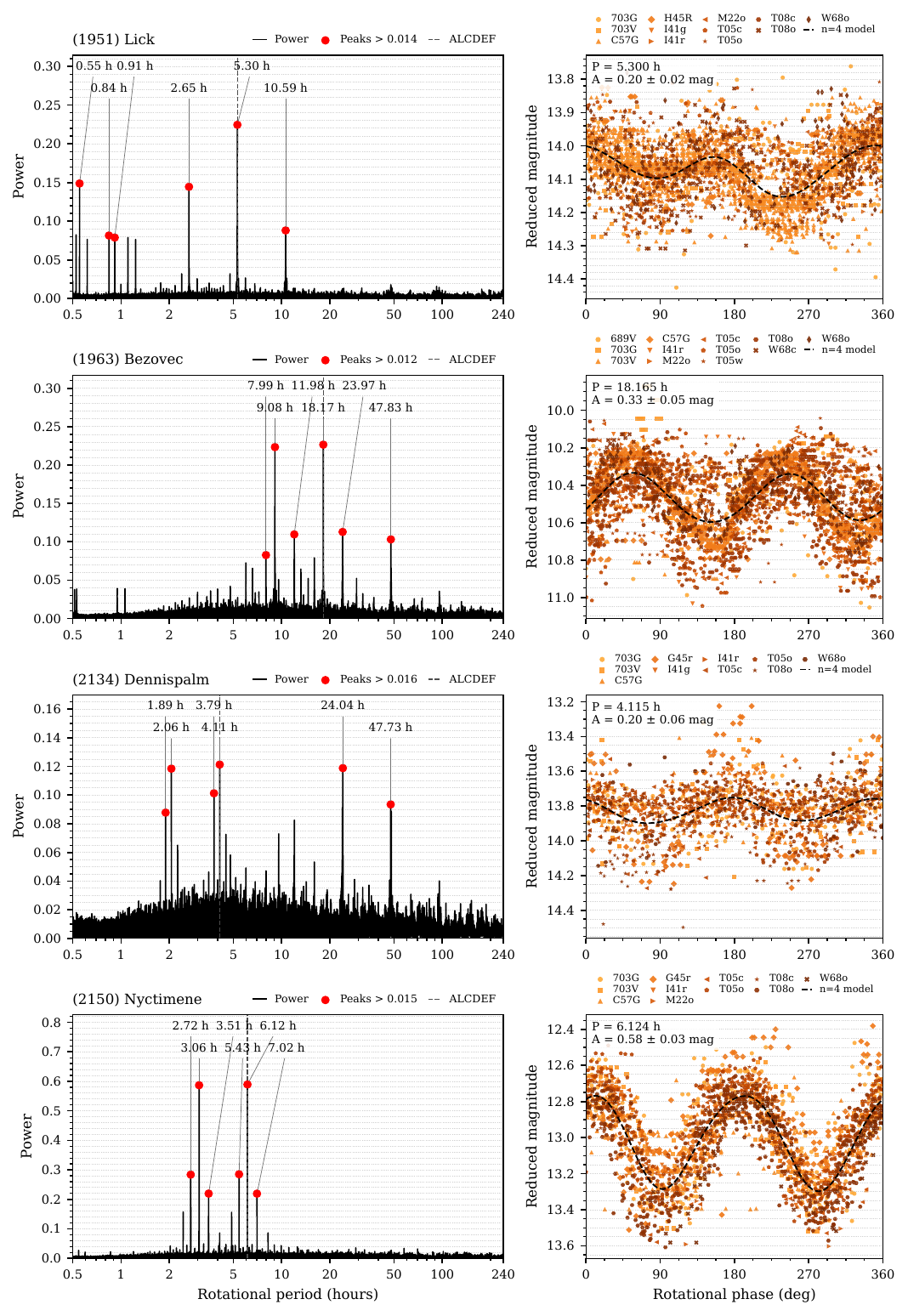}
    \caption{Validation of the C-method with the four asteroids of known period. Left: power spectra ($n=2$) over $0.5$-$240$ hours, with the ALCDEF period marked by the dashed line. Right: light curves folded with the C-method period, with the $n=4$ Fourier model (dashed line) and phase 0 at the epochs $t_0$ of Table~\ref{tab:combined}.}
    \label{fig: validation}
\end{figure}

The four asteroids 1951, 1963, 2134 and 2150 have rotation periods in the ALCDEF database \citep{Warner2024}. Their MPC data were analyzed with both methods in the same way as the eight target asteroids, with T08o as the reference observatory of the C-method and the datasets of Tables~\ref{tab:asteroids} and \ref{tab:combined}. The candidates of the S-method were the strongest peaks of the largest datasets and of the C-method, as for the other asteroids. Fig.~\ref{fig: validation} shows the power spectra and light curves of the C-method, and the periods are given in Table~\ref{tab:results}.

For 1951 (Lick), the S-method gives $(5.2999 \pm 0.0006)$ hours from eight of the fourteen datasets, and the C-method gives $5.2995$ hours. They differ from the ALCDEF period of $5.302$ hours by $0.04\%$ and $0.05\%$. The C-method power spectrum has a clear peak at the period ($w = 0.22$) and weaker ones at its half ($w = 0.14$) and at $0.552$ hours ($w = 0.15$). The latter is the alias of the rotation frequency with the 30 minute sampling of the TESS data ($48 - 24/5.2995 \approx 24/0.552$ days$^{-1}$).

For 1963 (Bezovec), whose period is close to three quarters of a day, the C-method power spectrum has its strongest peaks at $18.165$ hours and its half ($w = 0.23$ and $0.22$). The window function peaks at $23.97$ and $47.8$ hours reach $w = 0.11$ and $0.10$. Without the TESS data, which have no daily gaps, these peaks were as strong as the period. The S-method gives $(18.163 \pm 0.001)$ hours from ten datasets.

For 2134 (Dennispalm), the period of $4.115$ hours has practically the same power ($w = 0.12$) as the window function peak at $24.04$ hours and as its own half. The S-method gives $(4.1144 \pm 0.0004)$ hours from five datasets, and the light curve folded with $4.115$ hours has two maxima and two minima.

For 2150 (Nyctimene), the large amplitude gives a strong peak at $6.124$ hours ($w = 0.59$), while the daily aliases at $5.43$ and $2.72$ hours reach only half of this power. Eleven of the thirteen datasets meet the criteria of the S-method, which gives $(6.1245 \pm 0.0002)$ hours, and the 704V and T05w periods are not consistent with them.

The amplitudes of the four asteroids (Section~\ref{subsec: amplitude}), $(0.20 \pm 0.02)$, $(0.33 \pm 0.05)$, $(0.20 \pm 0.06)$ and $(0.58 \pm 0.03)$ mag, are within the ranges of the published amplitudes of the Lightcurve Database \citep{Warner2009}, which are $0.15$-$0.33$, $0.15$-$0.60$, $0.09$-$0.27$ and $0.56$-$0.90$ mag. The medians of the amplitudes from dense light curves in this database are $0.28$, $0.49$, $0.26$ and $0.62$ mag. The peak-to-peak ranges of the $n=4$ models of the combined data, $0.16$, $0.26$, $0.17$ and $0.54$ mag, are smaller, and for 2150 the value is below the published range.

\section{Long-period analysis of (3173) McNaught}\label{app: 3173}

\begin{figure}
    \centering
    \includegraphics[width=1\linewidth]{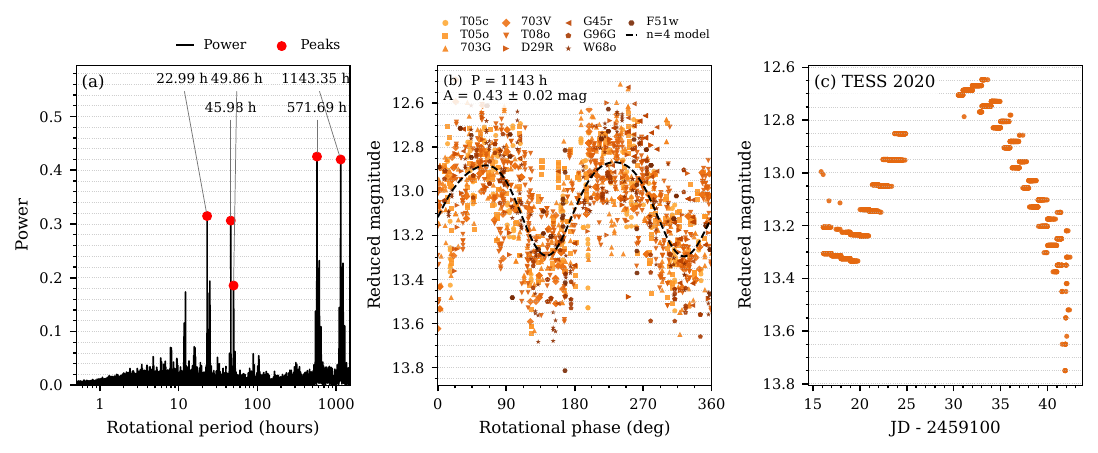}
    \caption{Long-period analysis of asteroid 3173. (a) Power spectrum ($n=2$) of the combined data over $0.5$-$1500$ hours. (b) Light curve folded with $1143$ hours with the $n=4$ Fourier model (dashed line). (c) TESS measurements of 2020 (C57G) against time.}
    \label{fig: 3173 long}
\end{figure}

The power spectrum of the combined data of 3173 (Section~\ref{subsec: 3173}) has its strongest peaks at $22.99$ and $45.98$ hours, with the daily alias at $49.9$ hours. To test whether these peaks come from a slower variation, the search of the C-method was extended to $1500$ hours (Fig.~\ref{fig: 3173 long}a). Two peaks dominate this range, at $572$ hours ($w = 0.43$) and $1143$ hours ($w = 0.42$). They correspond to a single-peaked and a double-peaked light curve of the same variation, which is shown folded with $1143$ hours in Fig.~\ref{fig: 3173 long}b. Its amplitude in the $n=4$ model is $(0.43 \pm 0.02)$ mag. The $46$ and $50$ hour peaks are daily aliases of this variation, as $48/45.98 - 48/1143 \approx 1$ days$^{-1}$ and $48/49.9 + 48/1143 \approx 1$ days$^{-1}$.

To estimate the probability of the candidates, the combined data were resampled 300 times with replacement, in blocks of one night of one dataset, and the candidate with the highest power was recorded each time. The $572$ hour period has the highest power in $84\%$ of the resamples and the $1143$ hour period in $16\%$. The short periods ($22.99$, $24.93$, $45.98$ and $49.9$ hours) never have the highest power. The false alarm probability of the long-period peak was estimated by permuting the nightly mean magnitudes between the nights, which keeps the sampling and the scatter within each night but destroys any slow variation. In 300 permutations, the highest power between $240$ and $1500$ hours had a median of $0.037$ and a 99th percentile of $0.054$. It never reached the observed $w = 0.43$, so the false alarm probability is below $0.004$.

The TESS data were excluded from the C-method by criterion (3) of step 7. Their magnitudes are constant over intervals of up to about one day, so they form streaks of equal brightness (Fig.~\ref{fig: 3173 long}c). The visit of 2020 covers two stretches of 9 and 12 days without daily gaps. The brightness changes slowly within each stretch, but no $46$ or $50$ hour variation larger than about $0.05$ mag is present. When the TESS data are added to the combined data, the $46$ and $50$ hour peaks almost disappear ($w \leq 0.05$). The strongest peaks of the 0.5-240 hour search are then at its upper end (236 hours, $w = 0.14$), and the $1143$ and $572$ hour peaks become stronger ($w = 0.61$ and $0.60$). The $1143$ hour period then has the highest power in $75\%$ of the bootstrap resamples. Adding also the I41r and M22o datasets gives the same result, with the $1143$ hour period in $83\%$ of the resamples. Therefore, the TESS data do not make the short period clearer, but they confirm the slow variation.

A rotation period of $572$ or $1143$ hours ($24$ or $48$ days) would make 3173 a very slow rotator. Offsets that change between nights or oppositions are one possible origin of a slow variation (Section~\ref{sec: three}). The TESS data of a single instrument also show a slow change, but blending with stars in the large TESS pixels can produce such changes as well. The choice between a single-peaked and a double-peaked light curve, the difference between the synodic and sidereal periods over the 21-year baseline, and a possible non-principal axis rotation cannot be resolved with the present data. For these reasons, no long rotation period is adopted in this work, and the rotation period of 3173 remains undetermined.

\clearpage
\bibliography{refs}

@article{Zappala1990,
  author  = {Zappal{\`a}, V. and Cellino, A. and Barucci, A. M. and Fulchignoni, M. and Lupishko, D. F.},
  title   = {An analysis of the amplitude-phase relationship among asteroids},
  journal = {Astronomy and Astrophysics},
  volume  = {231},
  pages   = {548--560},
  year    = {1990}
}

@article{Chapman1994,
  author  = {Chapman, C. R. and Morrison, D.},
  title   = {Impacts on the {E}arth by asteroids and comets: assessing the hazard},
  journal = {Nature},
  volume  = {367},
  pages   = {33--40},
  year    = {1994},
  doi     = {10.1038/367033a0}
}

@article{Shuvalov2013,
  author  = {Shuvalov, V. V. and Svettsov, V. V. and Trubetskaya, I. A.},
  title   = {An estimate for the size of the area of damage on the {E}arth's surface after impacts of 10--300-m asteroids},
  journal = {Solar System Research},
  volume  = {47},
  pages   = {260--267},
  year    = {2013},
  doi     = {10.1134/S0038094613040217}
}

@article{Waszczak2015,
  author  = {Waszczak, A. and Chang, C.-K. and Ofek, E. O. and others},
  title   = {Asteroid light curves from the {P}alomar {T}ransient {F}actory survey: rotation periods and phase functions from sparse photometry},
  journal = {The Astronomical Journal},
  volume  = {150},
  pages   = {75},
  year    = {2015},
  doi     = {10.1088/0004-6256/150/3/75}
}

@article{Chang2015,
  author  = {Chang, C.-K. and Ip, W.-H. and Lin, H.-W. and others},
  title   = {Asteroid spin-rate study using the {I}ntermediate {P}alomar {T}ransient {F}actory},
  journal = {The Astrophysical Journal Supplement Series},
  volume  = {219},
  pages   = {27},
  year    = {2015},
  doi     = {10.1088/0067-0049/219/2/27}
}

@article{Scargle1982,
  author  = {Scargle, J. D.},
  title   = {Studies in astronomical time series analysis. {II}. Statistical aspects of spectral analysis of unevenly spaced data},
  journal = {Astrophysical Journal},
  volume  = {263},
  year    = {1982},
  pages   = {835--853},
  doi     = {10.1086/160554}
}

@article{Lomb1976,
  author  = {Lomb, N. R.},
  title   = {Least-Squares Frequency Analysis of Unequally Spaced Data},
  journal = {Astrophysics and Space Science},
  volume  = {39},
  pages   = {447--462},
  year    = {1976},
  doi     = {10.1007/BF00648343}
}

@article{Veres2015,
  author  = {Vere{\v{s}}, P. and Jedicke, R. and Fitzsimmons, A. and Denneau, L. and Granvik, M. and Bolin, B. and Chastel, S. and Wainscoat, R. J. and Burgett, W. S. and Chambers, K. C. and Flewelling, H. and Kaiser, N. and Magnier, E. A. and Morgan, J. S. and Price, P. A. and Tonry, J. L. and Waters, C.},
  title   = {Absolute magnitudes and slope parameters for 250{,}000 asteroids observed by Pan-{STARRS} {PS1} -- Preliminary results},
  journal = {Icarus},
  volume  = {261},
  pages   = {34--47},
  year    = {2015},
  doi     = {10.1016/j.icarus.2015.08.007}
}

@article{Muinonen2010,
  author  = {Muinonen, K. and Belskaya, I. and Cellino, A. and Delb{\`o}, M. and Levasseur-Regourd, A. C. and Penttil{\"a}, A. and  Tedesco, E.},
    title   = {A three-parameter magnitude phase function for asteroids},
  journal = {Icarus},
  volume  = {209},
  pages   = {542-555},
  year    = {2010},
  doi     = {10.1016/j.icarus.2010.04.003}
}

@book{ivezic2020statistics,
  title={Statistics, data mining, and machine learning in astronomy: a practical {Python} guide for the analysis of survey data},
  author={Ivezi{\'c}, {\v{Z}}eljko and Connolly, Andrew J and VanderPlas, Jacob T and Gray, Alexander},
  volume={8},
  year={2020},
  publisher={Princeton University Press},
  doi={10.2307/j.ctvrxk1hs}
}

@article{bonanno2000analytical,
  title={An analytical approximation for the {MOID} and its consequences},
  author={Bonanno, Claudio},
  journal={Astronomy and Astrophysics},
  volume={360},
  pages={411--416},
  year={2000}
}

@incollection{Durech2015,
  author    = {Durech, J. and Carry, B. and Delb{\`o}, M. and Kaasalainen, M. and Viikinkoski, M.},
  title     = {Asteroid models from multiple data sources},
  booktitle = {Asteroids IV},
  editor    = {Michel, P. and DeMeo, F. E. and Bottke, W. F.},
  publisher = {University of Arizona Press},
  address   = {Tucson},
  pages     = {183--202},
  year      = {2015},
  doi       = {10.2458/azu_uapress_9780816532131-ch010}
}

@article{Warner2009,
  author  = {Warner, B. D. and Harris, A. W. and Pravec, P.},
  title   = {The asteroid lightcurve database},
  journal = {Icarus},
  volume  = {202},
  number  = {1},
  pages   = {134--146},
  year    = {2009},
  doi     = {10.1016/j.icarus.2009.02.003}
}

@article{Penttila2016,
  author  = {Penttil{\"a}, A. and Shevchenko, V. G. and Wilkman, O. and Muinonen, K.},
  title   = {{H}, {G}$_1$, {G}$_2$ photometric phase function extended to low-accuracy data},
  journal = {Planetary and Space Science},
  volume  = {123},
  pages   = {117--125},
  year    = {2016},
  doi     = {10.1016/j.pss.2015.08.010}
}

@article{AliLagoa2017,
  author  = {Al{\'i}-Lagoa, V. and Delbo', M.},
  title   = {Sizes and albedos of {Mars}-crossing asteroids from {WISE/NEOWISE} data},
  journal = {Astronomy and Astrophysics},
  volume  = {603},
  pages   = {A55},
  year    = {2017},
  doi     = {10.1051/0004-6361/201629917}
}

@article{Colazo2021,
  author  = {Colazo, M. and Duffard, R. and Weidmann, W.},
  title   = {The determination of asteroid {H} and {G} phase function parameters using {Gaia} {DR2}},
  journal = {Monthly Notices of the Royal Astronomical Society},
  volume  = {504},
  pages   = {761--768},
  year    = {2021},
  doi     = {10.1093/mnras/stab726}
}

@article{Eglitis2025,
  author  = {Eglitis, I. and Svincicka, D.},
  title   = {Rotation period estimates for 14 asteroids with the {Earth} {MOID} less than 1.1 {AU}},
  journal = {Icarus},
  volume  = {428},
  pages   = {116380},
  year    = {2025},
  doi     = {10.1016/j.icarus.2024.116380}
}

@article{Harris2014,
  author  = {Harris, A. W. and Pravec, P. and Gal{\'a}d, A. and others},
  title   = {On the maximum amplitude of harmonics of an asteroid lightcurve},
  journal = {Icarus},
  volume  = {235},
  pages   = {55--59},
  year    = {2014},
  doi     = {10.1016/j.icarus.2014.03.004}
}

@article{Hoffmann2025,
  author  = {Hoffmann, T. and Micheli, M. and Cano, J. L. and Devog{\`e}le, M. and Farnocchia, D. and Pravec, P. and Vere{\v{s}}, P. and Poppe, B.},
  title   = {Debiasing astro-photometric observations with corrections using statistics ({DePhOCUS})},
  journal = {Icarus},
  volume  = {426},
  pages   = {116366},
  year    = {2025},
  doi     = {10.1016/j.icarus.2024.116366}
}

@article{Kaasalainen2001,
  author  = {Kaasalainen, M. and Torppa, J. and Muinonen, K.},
  title   = {Optimization methods for asteroid lightcurve inversion: {II}. The complete inverse problem},
  journal = {Icarus},
  volume  = {153},
  pages   = {37--51},
  year    = {2001},
  doi     = {10.1006/icar.2001.6674}
}

@article{Mahlke2021,
  author  = {Mahlke, M. and Carry, B. and Denneau, L.},
  title   = {Asteroid phase curves from {ATLAS} dual-band photometry},
  journal = {Icarus},
  volume  = {354},
  pages   = {114094},
  year    = {2021},
  doi     = {10.1016/j.icarus.2020.114094}
}

@article{Mainzer2011,
  author  = {Mainzer, A. K. and Grav, T. and Masiero, J. and Hand, E. and Bauer, J. and Tholen, D. J. and McMillan, R. S. and Spahr, T. and Cutri, R. M. and Wright, E. L. and Watkins, J. and Mo, W. and Maleszewski, C.},
  title   = {{NEOWISE} Studies of Spectrophotometrically Classified Asteroids: Preliminary Results},
  journal = {The Astrophysical Journal},
  volume  = {741},
  pages   = {90},
  year    = {2011},
  doi     = {10.1088/0004-637X/741/2/90}
}

@misc{Mainzer2016,
  author  = {Mainzer, A. K. and Bauer, J. M. and Cutri, R. M. and Grav, T. and Kramer, E. A. and Masiero, J. R. and Nugent, C. R. and Sonnett, S. M. and Stevenson, R. A. and Wright, E. L.},
  title   = {{NEOWISE} Diameters and Albedos V1.0},
  howpublished = {NASA Planetary Data System},
  year    = {2016},
  doi     = {10.26033/ddgd-9m07},
  url     = {https://doi.org/10.26033/ddgd-9m07}
}

@misc{Mainzer2019,
  author  = {Mainzer, A. K. and Bauer, J. M. and Cutri, R. M. and Grav, T. and Kramer, E. A. and Masiero, J. R. and Sonnett, S. and Wright, E. L.},
  title   = {{NEOWISE} Diameters and Albedos V2.0},
  howpublished = {NASA Planetary Data System},
  year    = {2019},
  doi     = {10.26033/18S3-2Z54},
  url     = {https://doi.org/10.26033/18S3-2Z54}
}

@article{Nugent2015,
  author  = {Nugent, C. R. and Mainzer, A. and Masiero, J. and Bauer, J. and Cutri, R. M. and Grav, T. and Kramer, E. and Sonnett, S. and Stevenson, R. and Wright, E. L.},
  title   = {{NEOWISE} Reactivation Mission Year One: Preliminary Asteroid Diameters and Albedos},
  journal = {The Astrophysical Journal},
  volume  = {814},
  pages   = {117},
  year    = {2015},
  doi     = {10.1088/0004-637X/814/2/117}
}

@article{Nugent2016,
  author  = {Nugent, C. R. and Mainzer, A. and Bauer, J. and Cutri, R. M. and Kramer, E. A. and Grav, T. and Masiero, J. and Sonnett, S. and Wright, E. L.},
  title   = {{NEOWISE} Reactivation Mission Year Two: Asteroid Diameters and Albedos},
  journal = {The Astronomical Journal},
  volume  = {152},
  pages   = {63},
  year    = {2016},
  doi     = {10.3847/0004-6256/152/3/63}
}

@article{Pravec2007,
  author  = {Pravec, P. and Harris, A. W. and Warner, B. D.},
  title   = {{NEA} rotations and binaries},
  journal = {Proceedings of the International Astronomical Union},
  volume  = {236},
  pages   = {167--176},
  year    = {2007},
  doi     = {10.1017/S1743921307003201}
}

@article{VanderPlas2018,
  author  = {VanderPlas, J. T.},
  title   = {Understanding the {Lomb}--{Scargle} Periodogram},
  journal = {The Astrophysical Journal Supplement Series},
  volume  = {236},
  pages   = {16},
  year    = {2018},
  doi     = {10.3847/1538-4365/aab766}
}

@misc{Warner2024,
  author  = {Warner, B. D.},
  title   = {{ALCDEF} Light curve database},
  year    = {2024},
  howpublished = {\url{https://alcdef.org/}}
}

@article{Eglitis2023,
  author  = {Eglitis, I. and Nagainis, K.},
  title   = {Light curve analysis of main belt asteroids 4747, 5255, 11411, 15433, 17866},
  journal = {Open Astronomy},
  volume  = {32},
  pages   = {20220227},
  year    = {2023},
  doi     = {10.1515/astro-2022-0227}
}

@ARTICLE{astropy:2022,
       author = {{Astropy Collaboration} and {Price-Whelan}, A.~M. and {Lim}, P.~L. and {Earl}, N. and others},
        title = "{The Astropy Project: Sustaining and Growing a Community-oriented Open-source Project and the Latest Major Release (v5.0) of the Core Package}",
      journal = {The Astrophysical Journal},
         year = 2022,
        month = aug,
       volume = {935},
       number = {2},
          eid = {167},
        pages = {167},
          doi = {10.3847/1538-4357/ac7c74},
archivePrefix = {arXiv},
       eprint = {2206.14220},
 primaryClass = {astro-ph.IM},
       adsurl = {https://ui.adsabs.harvard.edu/abs/2022ApJ...935..167A}
}

@article{Carry2016,
  author  = {Carry, B. and Solano, E. and Eggl, S. and DeMeo, F. E.},
  title   = {Spectral properties of near-{Earth} and {Mars}-crossing asteroids using {Sloan} photometry},
  journal = {Icarus},
  volume  = {268},
  pages   = {340--354},
  year    = {2016},
  doi     = {10.1016/j.icarus.2015.12.047}
}

@article{Slyusarev2025,
  author  = {Slyusarev, I. G. and Shevchenko, V. G. and Belskaya, I. N. and Krugly, Yu. N. and others},
  title   = {Opposition effect of {M}-type asteroids},
  journal = {Planetary and Space Science},
  volume  = {260},
  pages   = {106103},
  year    = {2025},
  doi     = {10.1016/j.pss.2025.106103}
}

@article{Oszkiewicz2023,
  author  = {Oszkiewicz, D. and Troianskyi, V. and Gal{\'a}d, A. and others},
  title   = {Spins and shapes of basaltic asteroids and the missing mantle problem},
  journal = {Icarus},
  volume  = {397},
  pages   = {115520},
  year    = {2023},
  doi     = {10.1016/j.icarus.2023.115520}
}

@article{Durech2024,
  author  = {Durech, J. and Vokrouhlick{\'y}, D. and Pravec, P. and others},
  title   = {Secular change in the spin states of asteroids due to radiation and gravitation torques. {N}ew detections and updates of the {YORP} effect},
  journal = {Astronomy \& Astrophysics},
  volume  = {682},
  pages   = {A93},
  year    = {2024},
  doi     = {10.1051/0004-6361/202348350}
}

@article{Colazo2025,
  author  = {Colazo, M. and Oszkiewicz, D. and Alvarez-Candal, A. and Po{\.z}niak, P. and Bartczak, P. and Podlewska-Gaca, E.},
  title   = {Asteroid phase curves and phase coloring effect using the {ATLAS} survey data},
  journal = {Icarus},
  volume  = {436},
  pages   = {116577},
  year    = {2025},
  doi     = {10.1016/j.icarus.2025.116577}
}

@article{Rommel2025,
  author  = {Rommel, F. L. and Fern{\'a}ndez-Valenzuela, E. and Souza-Feliciano, A. C. and Morgado, B. E. and Proudfoot, B.},
  title   = {{SpinTrace}: A {P}ython library to study the physical properties of small bodies using survey-based photometry},
  journal = {The Planetary Science Journal},
  volume  = {6},
  pages   = {280},
  year    = {2025},
  doi     = {10.3847/PSJ/ae1c3a}
}

@article{Xu2025,
  author  = {Xu, X. and Wang, X. and Muinonen, K. and Gu, S. and Penttil{\"a}, A. and Xu, F. and Sun, L. and Huang, J. and Zhang, P. and Wang, A.},
  title   = {Photometric analysis of asteroids in the {P}hocaea region},
  journal = {The Astronomical Journal},
  volume  = {170},
  pages   = {35},
  year    = {2025},
  doi     = {10.3847/1538-3881/add3f4}
}

@incollection{Bowell1994,
  author    = {Bowell, E. and Muinonen, K.},
  title     = {{Earth}-crossing asteroids and comets: groundbased search strategies},
  booktitle = {Hazards Due to Comets and Asteroids},
  editor    = {Gehrels, T. and Matthews, M. S. and Schumann, A. M.},
  publisher = {University of Arizona Press},
  address   = {Tucson},
  pages     = {149--197},
  year      = {1994},
  doi       = {10.2307/j.ctv23khmpv.10}
}

@article{sergeyev2025rotation,
  title={Rotation periods of asteroids serendipitously observed by the {NASA}/{Kepler} {K2} mission},
  author={Sergeyev, AV and Carry, B and Eggl, S and Berthier, J and Santerne, A and Vachier, F and Shevchenko, VG},
  journal={Astronomy \& Astrophysics},
  volume={703},
  pages={A302},
  year={2025},
  publisher={EDP Sciences},
  doi={10.1051/0004-6361/202554052}
}

\end{document}